\documentclass[fleqn,usenatbib]{rasti}

\usepackage{newtxtext,newtxmath}
\usepackage[T1]{fontenc}

\DeclareRobustCommand{\VAN}[3]{#2}
\let\VANthebibliography\thebibliography
\def\thebibliography{\DeclareRobustCommand{\VAN}[3]{##3}\VANthebibliography}

\usepackage{graphicx}	% Including figure files
\usepackage{float}
\usepackage{amsmath}	% Advanced maths commands
\usepackage[utf8]{inputenc}
\usepackage{caption}
\usepackage{subcaption}
\usepackage{hyperref}
\usepackage{longtable}
\usepackage{booktabs}
\usepackage{textcase}
\usepackage[pagewise]{lineno}
\title[KSIM: simulating KIDSpec, a MKID spectrograph for the optical/NIR]{KSIM: simulating KIDSpec, a Microwave Kinetic Inductance Detector spectrograph for the optical/NIR}

\author[V. B. Hofmann et al.]{
V. Benedict Hofmann,$^{1}$\thanks{E-mail: benedict.hofmann@durham.ac.uk (VBH)}
Kieran O'Brien,$^{1}$
\\
$^{1}$Centre for Advanced Instrumentation, Department of Physics, Durham University, South Road, Durham, DH1 3LE, UK\\
}

\date{Accepted 26/05/2023. Received 23/05/2023; in original form 12/08/2022}

\pubyear{2022}

\begin{document}
\label{firstpage}
\pagerange{\pageref{firstpage}--\pageref{lastpage}}
\maketitle

\begin{abstract}
KIDSpec, the Kinetic Inductance Detector Spectrometer, is a proposed optical to near IR Microwave Kinetic Inductance Detector (MKID) spectrograph. MKIDs are superconducting photon counting detectors which are able to resolve the energy of incoming photons and their time of arrival. KIDSpec will use these detectors to separate incoming spectral orders from a grating, thereby not requiring a cross-disperser. In this paper we present a simulation tool for KIDSpec's potential performance upon construction to optimise a given design. This simulation tool is the KIDSpec Simulator (KSIM), a Python package designed to simulate a variety of KIDSpec and observation parameters. A range of astrophysical objects are simulated: stellar objects, an SDSS observed galaxy, a Seyfert galaxy, and a mock galaxy spectrum from the JAGUAR catalogue. Multiple medium spectral resolution designs for KIDSpec are simulated. The possible impact of MKID energy resolution variance and dead pixels were simulated, with impacts to KIDSpec performance observed using the Reduced Chi-Squared (RCS) value. Using dead pixel percentages from current instruments, the RCS result was found to only increase to 1.21 at worst for one of the designs simulated. SNR comparisons of object simulations between KSIM and X-Shooter's ETC were also simulated. KIDSpec offers a particular improvement over X-Shooter for short and faint observations. For a Seyfert galaxy ($m_{R}=21$) simulation with a 180s exposure, KIDSpec had an average SNR of 4.8, in contrast to 1.5 for X-Shooter. Using KSIM the design of KIDSpec can be optimised to improve the instrument further.
\end{abstract}


%%%%%%%%%%%%%%%%%%%%%%%%%%%%%%%%%%%%%%%%%%%%%%%%%%

%%%%%%%%%%%%%%%%% BODY OF PAPER %%%%%%%%%%%%%%%%%%

\section{Introduction}\label{intro}

Microwave Kinetic Inductance Detector (MKID) usage in optical/near-infrared (NIR) and sub-mm astronomy is growing. At the time of writing this paper, MKID's unique properties have been exploited in an increasing number of optical/NIR instruments: the Array Camera for Optical to Near-IR Spectrophotometry \citep[ARCONS;][]{Mazin2013}, the DARK-speckle Near-infrared Energy-resolving Superconducting Spectrophotometer \citep[DARKNESS;][]{Meeker2018}, and the MKID Exoplanet Camera \citep[MEC;][]{Walter2020}; with more instruments planned to take advantage of these superconducting detectors. One such instrument is the Kinetic Inductance Detector Spectrometer (KIDSpec), a conceptual medium resolution optical through near-IR MKID echelle spectrograph \citep{OBrien2020}. Here the MKIDs will allow for medium resolution spectroscopy instead of the spectro-imaging application of previous instruments.

MKID technology presents exciting opportunities for many areas of astronomy, especially with low-SNR spectroscopy remaining in large demand. MKIDs are well suited for this owing to an absence of read noise and dark current \citep{Mazin2018}, and inherent energy and microsecond time resolving capabilities. MKID's improvements to SNR can be seen from the standard SNR equation,

\begin{equation}
    SNR = \frac{N_{Obj}}{\sqrt{N_{Obj} + N_{Sky} + RON^2 + N_{Dark}}}
    \label{eqn: snr}
\end{equation}

where $N_{Obj}$ is the target object signal, $N_{Sky}$ is the sky background signal, $RON$ is the readout noise contribution, and $N_{Dark}$ is the dark current contribution. The lack of read noise and dark current components for MKIDs therefore allow for higher SNR at shorter exposures compared to typical semiconductor detectors. MKIDs are also more flexible, allowing rebinning to lower spectral resolution post-observation, without the additional readout noise contribution.

MKIDs can be multiplexed more easily than other superconducting detectors, such as Transition Edge Sensors (TESs) or Superconducting Tunnel Junctions (STJs). The largest MKID array currently in use exists on MEC, which consists of 20,440 MKIDs \citep{Walter2020}. 

KIDSpec will provide low-noise spectroscopy, as discussed above, for science involving faint sources such as high redshift galaxies. Examples of these faint sources are explored more in Sec. \ref{sdss_gal} and \ref{sey_gal}. These sources provide information on the stages of evolution of the universe, and studies of stellar content of these galaxies. MKID's lack of read noise and dark current will support these studies, alongside the device's time resolution allowing for other features such as superb cosmic ray removal. The time resolving capabilities of the MKIDs will also allow for observations of short period binary systems, such as eclipsing double white dwarf binaries, shown in \cite{Burdge2019}. These aspects of the MKID also allow for improved sky subtraction, owing to each spaxel not observing the object taking real time simultaneous sky background data \citep{Mazin2010}.

\begin{figure}
\centering
\includegraphics[width=9cm]{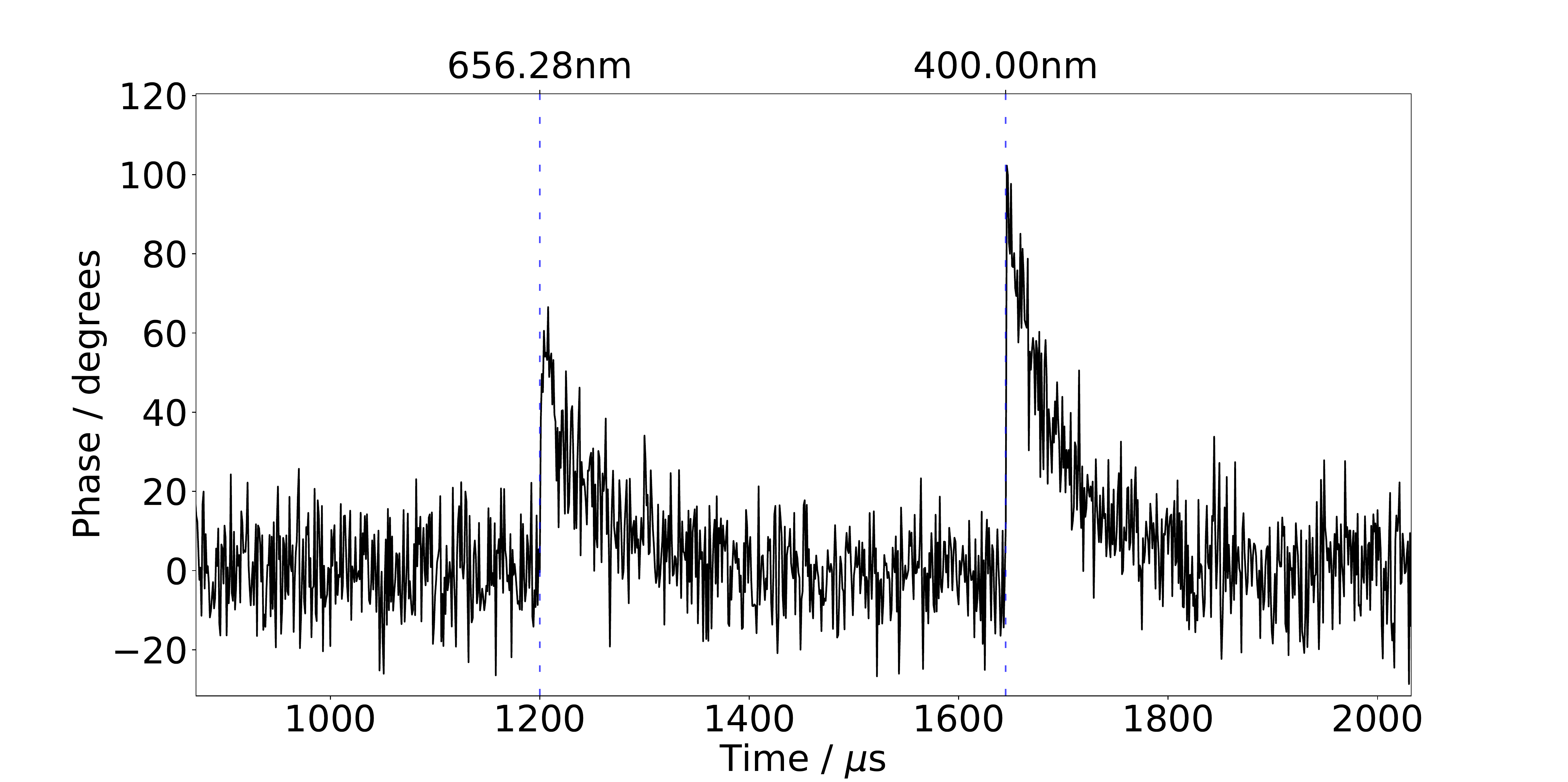}
    \caption{Simulated pulses in the phase time stream of an MKID pixel as the result of 656.28nm and 400.00nm photons, the H$\alpha$ wavelength and typical fudicial wavelength respectively. The height of these pulses is the maximum phase change (in degrees) which has been measured by the microwave probe signal. The right pulse has been caused by a 400.00nm photon and the left by a 656.28nm photon.}
    \label{fig: pulse_example}
\end{figure}
%KIDSpec's benefits, outlined in more detail in subsequent sections, can make optimum use of telescopes such as this.
%One such science area where KIDSpec could contribute where CCD detectors cannot as easily is short period binary systems. These systems can have periods of the order of 7 minutes \citep{Burdge2019GeneralSystem}, and while CCDs can currently observe these systems as seen in \cite{Burdge2019GeneralSystem}, KIDSpec could do so with much shorter exposure times allowing for finer resolution of the system's period. This would lead to more constrained parameters for the system, improving on the large error range shown in \cite{Burdge2019GeneralSystem}. These better constrained parameters would allow for more accurate tests of general relativity especially ahead of the launch of the Laser Interferometer Space Antenna (LISA) which will observe many of these systems for gravitational waves \citep{Amaro-Seoane2017LaserProposal}

%KIDSpec will utilise the native energy resolution ($R_{E}$) of the photon counting MKIDs for order separation of incoming photons, thereby not requiring a cross disperser.

%-------------------------------------------------------------------
\section{Design}\label{design}

\subsection{Microwave Kinetic Inductance Detectors}\label{MKIDS}
MKIDs are superconducting detectors. They consist of a capacitor and an inductor making a resonant circuit. In order to avoid thermal excitations they must be operated below the critical temperature, at temperatures of $T\approx{10^{-1}}$K, thereby requiring a cryogenic refrigerator. This can be either a dilution refrigerator or an adiabatic demagnetisation refrigerator. Aluminium MKID devices have been operated at 120mK \citep{DeVisser2021}, and PtSi devices have been operated at a temperature of 100mK \citep{Meeker2018}. Once cooled, a signal can be passed through the resonant circuit. In the case of an MKID this signal is at microwave frequencies.

%The current through the system is carried by Cooper pairs, which are pairs of loosely bound electrons bonded via the electron phonon interaction. When an AC electric field is applied near the surface of the superconductor it causes the Cooper pairs to accelerate, applying kinetic energy to these pairs \citep{Day2003AArrays}.
%This results in extra inductance caused by the extraction of the kinetic energy of the Cooper pairs. To do this the field direction must be reversed. Measurement of this can be done by placing the superconductor within a lithographed resonator \citep{Mazin2012AAstrophysics}. This change in inductance then causes a change in impedance. In the case of an incoming photon, when absorbed it will break a number of Cooper pairs depending on its energy. As a result of this a cascade of interacting quasiparticles is formed, which causes a change in surface impedance. The effects of an incoming photon onto the KID combine to alter the amplitude and phase of the microwave signal transferred through the circuit \citep{Mazin2004MicrowaveDetectors}.

When observing the microwave signal phase with respect to time,  an incoming photon will appear as a fast rise followed by an exponential decay, as shown in Fig. \ref{fig: pulse_example}. This fast rise occurs because of the photon breaking Cooper pairs as it strikes the MKID, causing a change in inductance and subsequently, a change in the phase of the microwave signal. These pairs then recombine with a characteristic timescale on the order of 10s of microseconds. 

The quantum efficiency (QE) of MKIDs has been measured to be between 0.73 and 0.22 for the range 200 to 3000nm respectively \citep{Mazin2010}. However, more recent work has begun to improve this to >80\% absorption from 400 to 1500nm \citep{Kouwenhoven2021}, using a method shown in \cite{Dai2019MeasurementDetectors} where efficiencies $\geq 90\%$ were demonstrated at 1550nm. 

A higher energy photon will break more Cooper pairs in the MKID on arrival, which then causes a greater phase change, and as a result,  MKIDs are capable of detecting the energy of incoming photons from the height of the pulses. 

The phase time stream in Fig. \ref{fig: pulse_example} also demonstrates the time resolving capability of MKIDs, where a photon's arrival time to $\approx1\mu$s can be measured from the leading edge of the pulse. 

Because of the shape of a photon event, dark excitations can be separated from photon excitations, and MKIDs do not suffer from read noise \citep{Mazin2018}. As with other detectors, the fabrication of MKIDs may have errors, primarily the dead pixel fraction (Walter et al. 2020), which could affect observations. This is explored more in Sec. \ref{mkid_eff}. More details on MKIDs, their operation, and the current state of the technology can be found in \cite{Day2003}, \cite{Mazin2004}, \cite{Mazin2013}, \cite{Mazin2019}, \cite{DeVisser2021}, and \cite{Zobrist2022}.

\subsection{KIDSpec design}\label{kidspec_design}

\begin{figure}
\includegraphics[width=9cm]{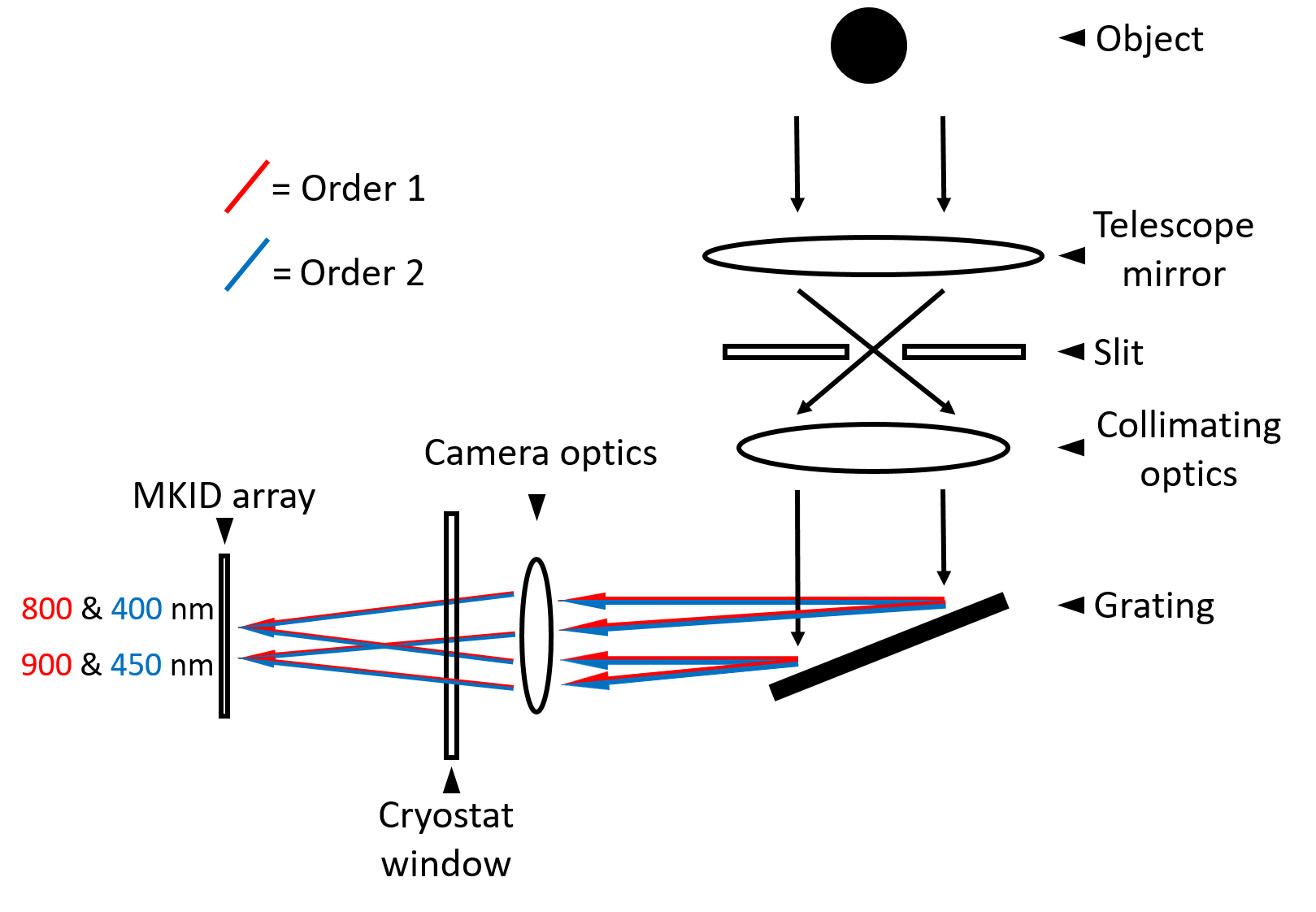}
    \caption{Conceptual optical layout of KIDSpec. Note that a cross disperser is not required. Shown after the grating are the first and second orders, arbitrarily chosen to show an example of separate order wavelengths which are incident on the MKIDs. Each set of wavelengths from the orders shown are exposed onto a single MKID, which can then separate the different orders. In practice many orders, and hence wavelengths, would be exposed onto a single MKID.}
    \label{fig: optical_layout}
\end{figure}

\begin{figure}
\includegraphics[width=\columnwidth]{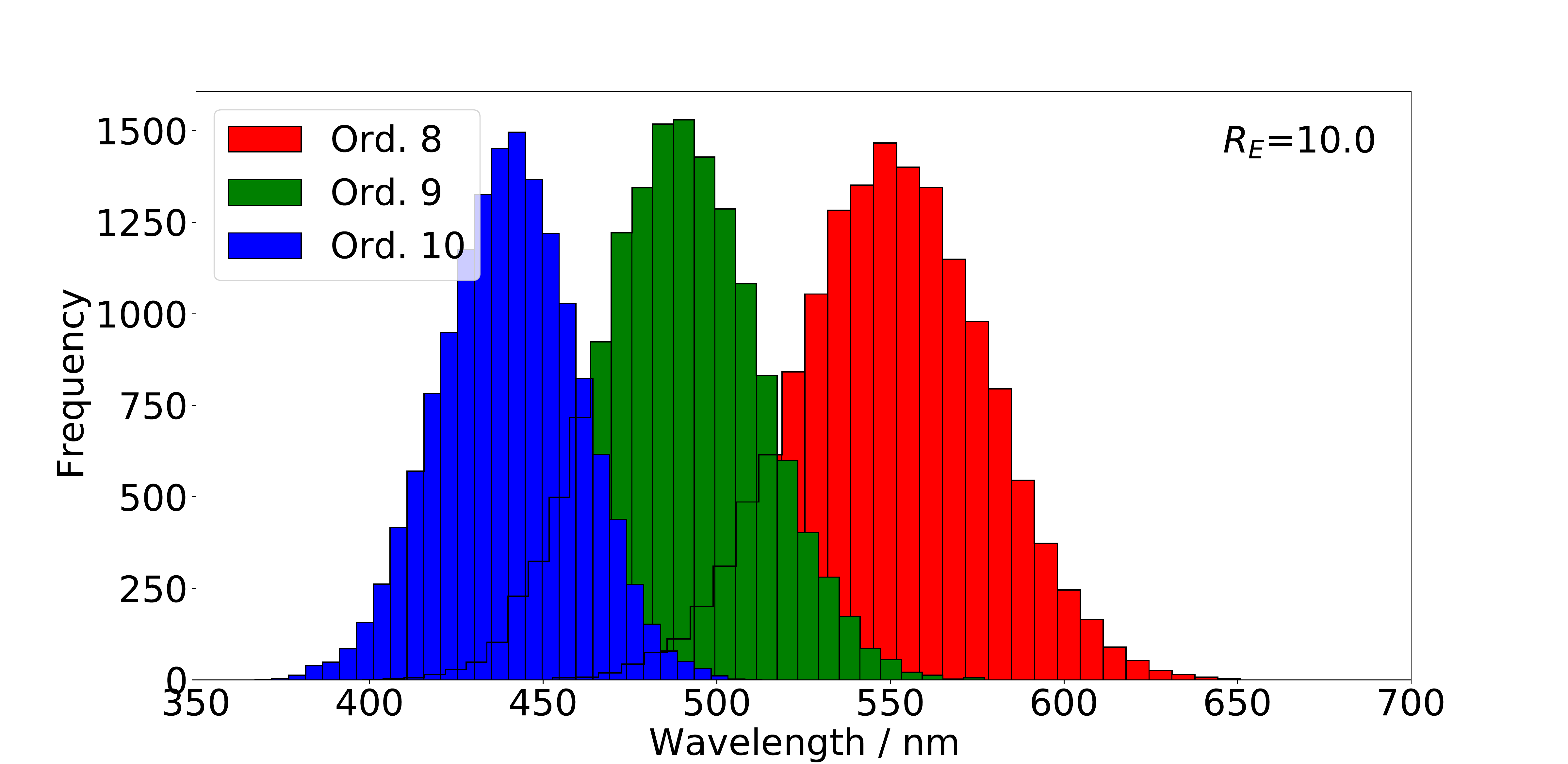}
\includegraphics[width=\columnwidth]{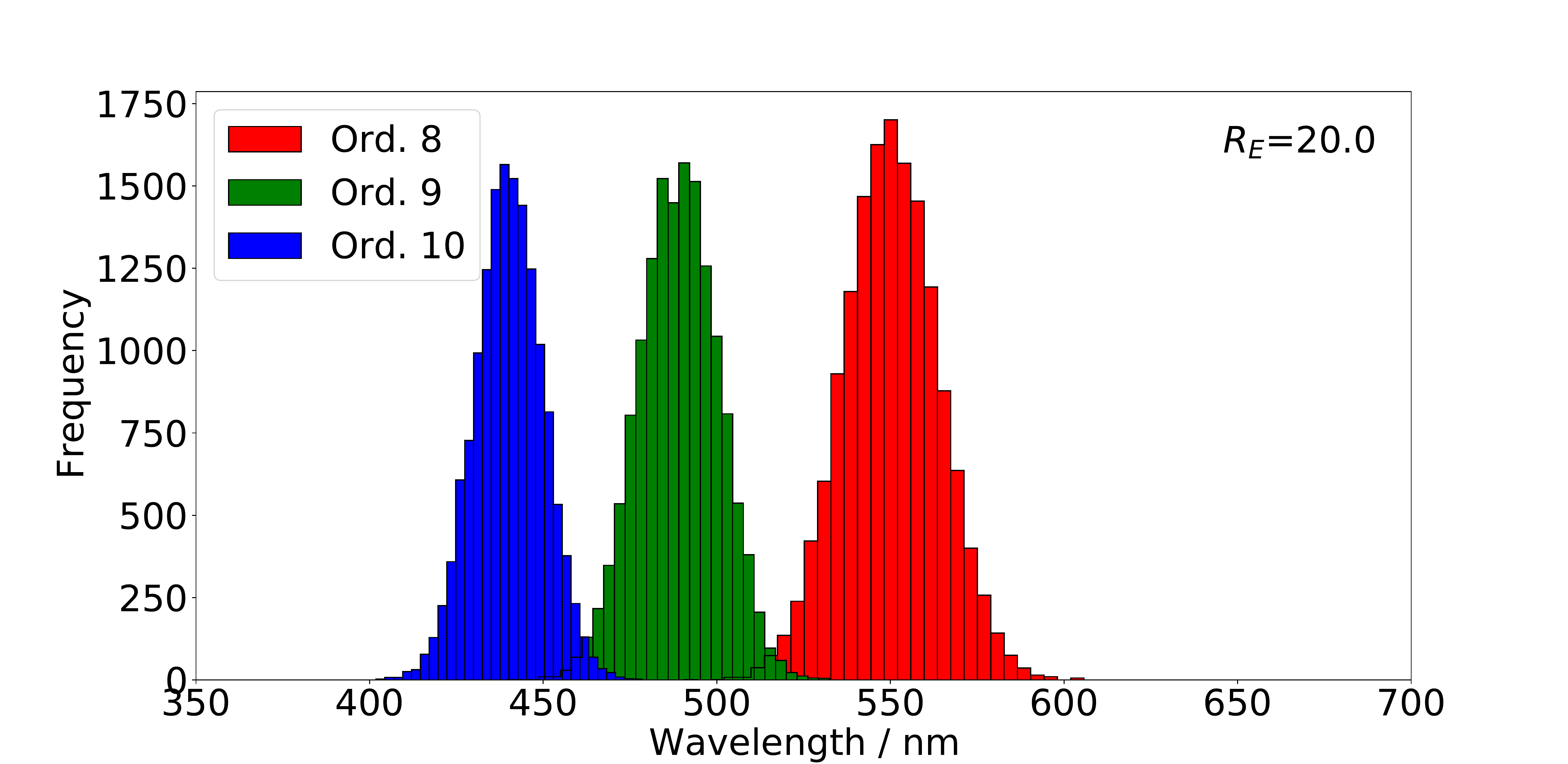}
\includegraphics[width=\columnwidth]{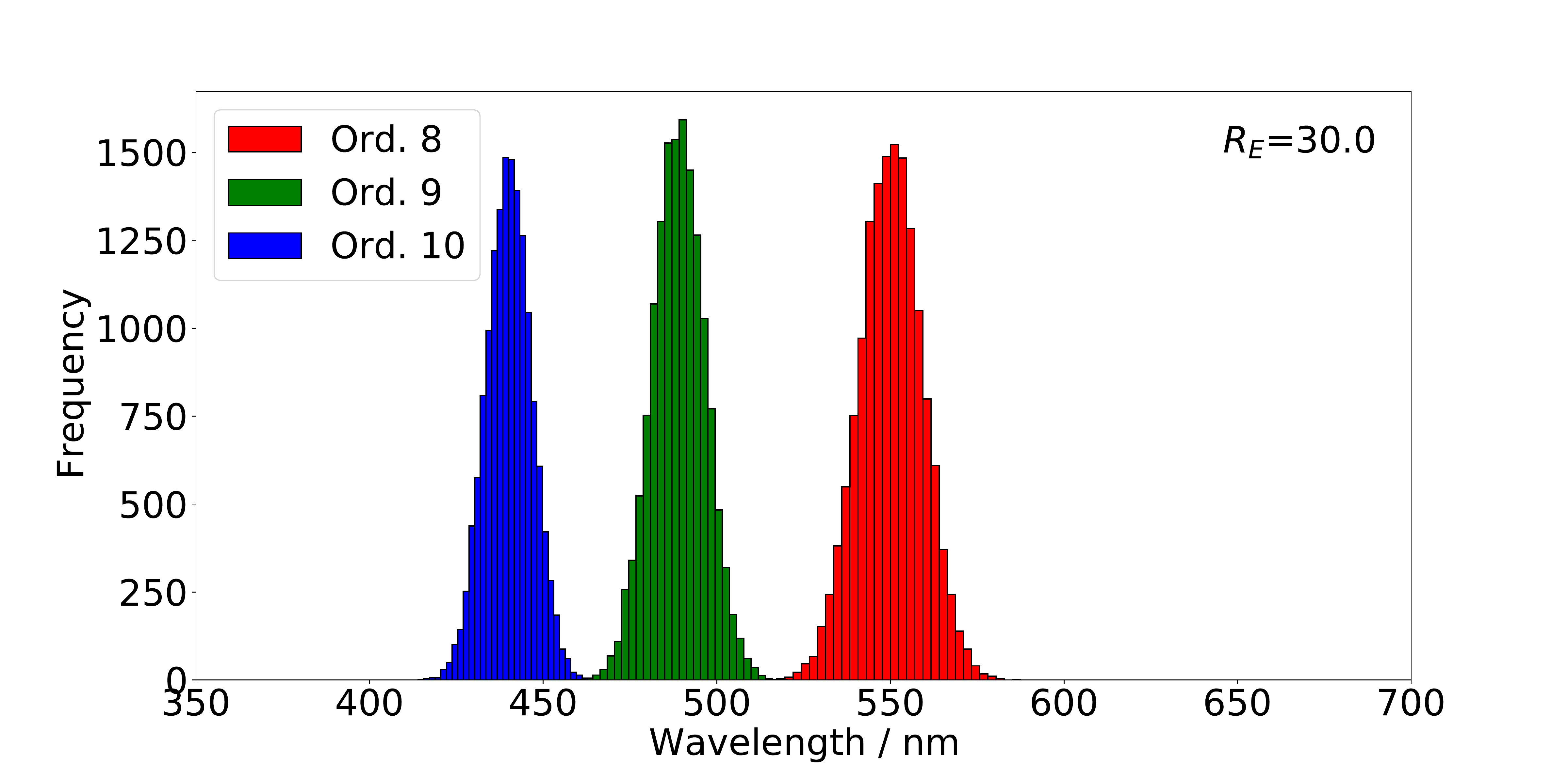}
    \caption{Three adjacent order Gaussians for orders 8, 9, and 10 with wavelengths 550, 489, and 440 nm at varying $R_{E}$. Values used here for $R_{E}$ are 10, 20, and 30, which are equal to the order number $m$, $2m$, and $3m$, where $m = 10$ is the order number. Note that if the $R_{E}$ was higher, the MKID could resolve higher order numbers. The Gaussians were generated using the photon time stream simulation outlined in Sec. \ref{PTS} and exhibit the spread of wavelengths of the incoming photons as seen by the MKID because of its $R_{E}$. The percentage of photons falling in an overlapping region for an $R_{E}$ of 10, 20, and 30 were approximately 97\%, 38\%, and 0\% respectively.}
    \label{fig: order_resolving_energy_res}

\end{figure}

KIDSpec will be a single object spectrograph with sufficient spectral resolution to resolve background sky lines for their subtraction, with a bandpass of $0.35-1.8\mu m$ in this work, but observing the K band would also be possible \citep{OBrien2020}. The use of MKIDs will allow KIDSpec to photon count with $\mu s$ time resolution, without a cross disperser, described below. These features and the MKIDs low-noise capabilities makes KIDSpec an exciting instrument for many science cases, especially those which require short and/or faint exposures with a wide bandpass. Fig. \ref{fig: optical_layout} shows a conceptual optical layout for KIDSpec. This design is based on a slit spectrograph where each spatial resolution element is dispersed onto a linear array of MKIDs. The slit could be formed by a mask, or a fibre, or even an integral field unit (IFU). These would require additional MKID pixels but would deliver additional information such as a simultaneous sky measurement or spatial information on the source. We have chosen to keep the geometry here as simple as possible but more complex spaxel geometries could also be simulated in a similar manner.

A key parameter of MKIDs is how accurately they can determine the energy of a photon, their so-called energy resolution ($R_{E}$). There are a number of intrinsic and external factors that influence this, such as the amplifier noise and the generation recombination noise, which can be seen as phase noise in Fig. \ref{fig: pulse_example} \citep{Mazin2013} \citep{Yates2011}. Because of these factors, if a monochromatic source is exposed onto an MKID the pulse heights will not all be equal. The $R_{E}$ value of an MKID can be determined by the full width half maximum (FWHM) of the phase heights distribution, when the MKID has been exposed to monochromatic light \citep{Meeker2015}. This phase height distribution can be approximated as a Gaussian-like trend, as shown in Fig. \ref{fig: order_resolving_energy_res}. This $R_{E}$ allows KIDSpec to discern the incoming orders from the echelle grating, thereby not requiring a cross-disperser. The abscence of a cross-disperser will improve KIDSpec's throughput, owing to the instrument requiring fewer optical surfaces. The highest $R_{E}$ achieved for MKIDs in the optical/NIR regime is 55 at 402nm \citep{deVisser2020}. This is sufficient for KIDSpec but is still 2-3 times lower than the theoretical maximum for MKID energy resolution \citep{Mazin2010}.

To separate the incoming orders, each MKID pixel within KIDSpec would see a particular wavelength bin from each order exposed onto it, depending on the pixel's position in the linear array. The $R_{E}$ then determines the number of orders which can be separated, since a higher $R_{E}$ results in a smaller FWHM for the phase height Gaussians of each order, as seen in Fig. \ref{fig: order_resolving_energy_res}. Essentially, the higher the $R_{E}$ is of an MKID, the more orders can be resolved by the device. Conversely, as the $R_{E}$ lowers, the order Gaussians overlap more, and in a wider wavelength range it becomes unclear to which order an individual photon belongs, and it could be misidentified. From the top panel of Fig. \ref{fig: order_resolving_energy_res}, if a photon event appears as $\approx$475nm to the MKID, there is a finite probability that it could belong to any of the orders shown. In the bottom panel of Fig. \ref{fig: order_resolving_energy_res}, a $3\sigma$ separation between the orders is shown. For equal height Gaussians, the orders have a $3\sigma$ separation where $R_{E} \geq 3m$, where $m$ is the order number. The $R_{E}$ of the MKIDs in an array hence defines the highest grating order which can be resolved with a $3\sigma$ separation. However, by using a $2\sigma$ separation instead, this would reduce the number of MKIDs required for the same spectral resolution. The lower separation requirement means a higher maximum grating order. However, this increases the overlap between order Gaussians as seen in Fig. \ref{fig: order_resolving_energy_res}. To prevent photon misidentification issues, these Gaussians could be fitted to determine what photons are sorted into which order. However this does complicate the analysis of incoming photons, and also requires enough incoming photons to create the fits for the orders. For simplicity hereafter, a $3\sigma$ separation is used. 

A fabrication effect potentially affecting observations is $R_{E}$ variance ($R_{var}$). When an MKID array is fabricated, the $R_E$ of the MKID pixels may not all be equal to each other. \cite{Meeker2018} showed that in an array of MKIDs fabricated for DARKNESS, the $R_{E}$s had a normal distribution with a FWHM of $\approx{3}$. This variance in the $R_E$ of the MKIDs will affect how well each individual MKID can separate incoming wavelengths, meaning the width of the Gaussians shown in Fig. \ref{fig: order_resolving_energy_res} will vary between MKID pixels in the array. This fabrication effect in addition to the dead pixel fraction are explored in Sec. \ref{mkid_eff}.

KIDSpec is still in the conceptual phase but will distinguish itself from other spectrographs through the use of MKIDs. This will allow KIDSpec to be a photon counting, low-noise, high efficiency instrument. More details on KIDSpec can be found in \cite{OBrien2020}. 

To develop the design and establish the limitations of KIDSpec, a simulation tool has been created to test various design and observational parameters, for a range of science cases. This is the KIDSpec Simulator (KSIM). For the remainder of this paper we outline KSIM and its current features, and include simulations of a selection of KIDSpec's science cases to showcase the instrument's potential, and limitations, upon construction.

\section{KIDSpec Simulator}\label{simulation}

\begin{figure}
\includegraphics[width=8.5cm]{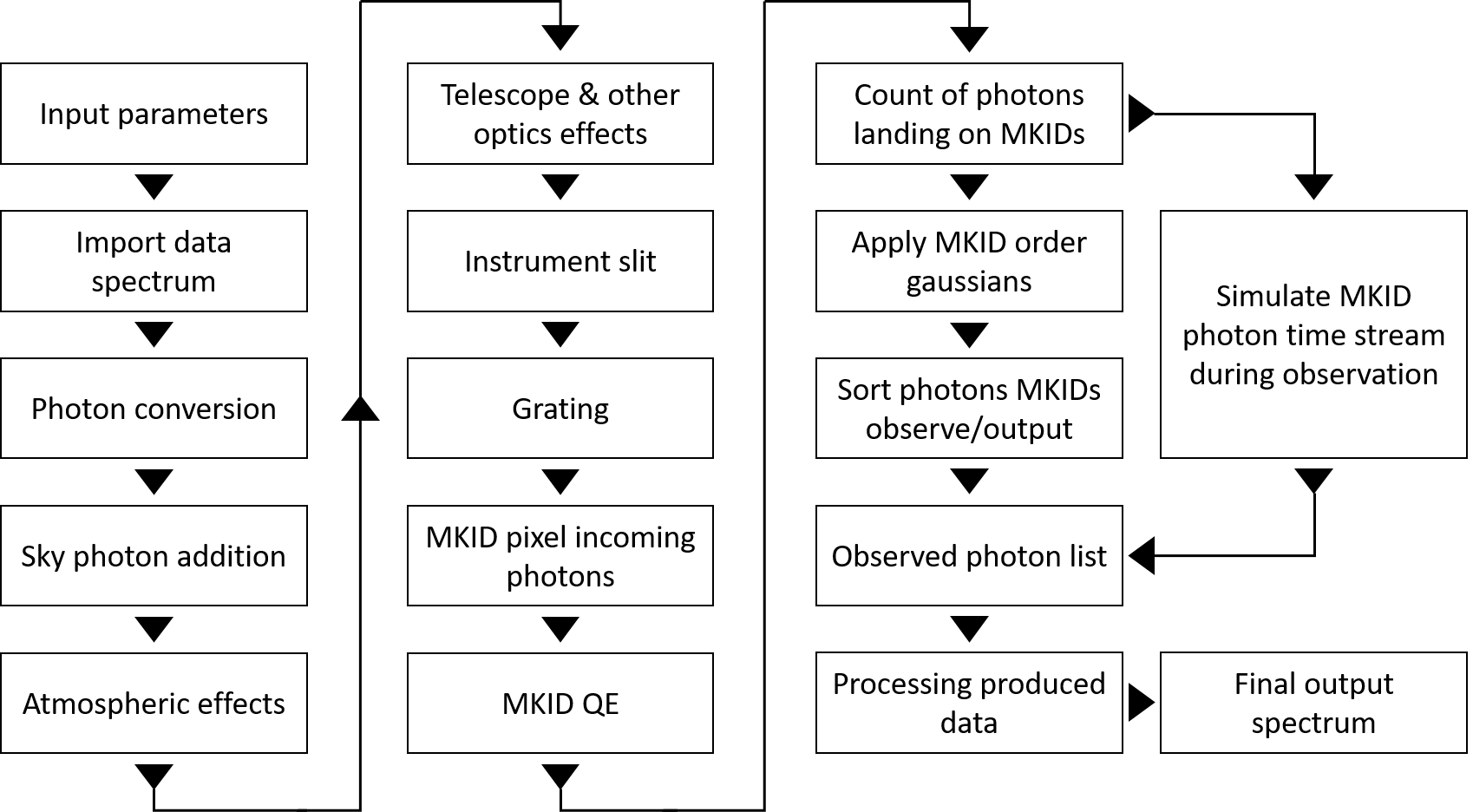}
    \caption{Flowchart depicting an overview of KSIM during the simulation of an object spectrum. The split in the directional arrows in the top right corner of the diagram indicates a choice of two methods for processing the MKID response to the incoming photons, the order Gaussians and photon time stream method, more details in Sec. \ref{pix_gaussian} and \ref{PTS}. }
    \label{fig: diagram_design}
\end{figure}

KSIM aims to give a performance prediction of what KIDSpec could achieve on sky. A key motivation for KSIM is flexibility, to be able to simulate a range of potential KIDSpec instruments, to give an overview simulation of what science those potential instruments could be used for. Fig. \ref{fig: diagram_design} shows a flowchart of KSIM. A parameter text file is read in, which includes aspects such as the grating parameters, telescope information, and atmospheric parameters. The full list of KSIM parameters are included in App. \ref{append_a}. All steps in KSIM are modular and so can be modified easily. Within KSIM, spectral pixels is defined as the number of MKID pixels used to make up an individual 1D spectrum. The effective number of MKID pixels is the total MKID pixels multiplied by the number of orders each MKID observes, more details in Sec. \ref{Grating} and \ref{pix_gaussian}.

The simulation initially reads in the data spectrum of the object to be simulated. As the simulation progresses, various processes are accounted for such as atmosphere and telescope transmission.

\subsection{Atmospheric effects, photon conversion, and telescope effects}\label{atmos_photon_tele}
Transmission data for the atmosphere was acquired from the GEMINI Observatory\footnote{\url{https://www.gemini.edu/}} in Cerro Pachon, Chile. The optical (OPT) extinction, and the near-infrared (NIR) atmospheric transmission\footnote{\url{https://www.gemini.edu/observing/telescopes-and-sites/sites\#Transmission}} data were utilised. The OPT extinction data was converted to the OPT transmission of the atmosphere using Eq. \ref{eqn: extinction_conv}, 

%\begin{equation}
%    \frac{I_2}{I_1} = \text{OPT}_{\text{Transmission}} = \exp\left(-\frac{A_{\text{atmos}}x}{2.5}\right) 
%    \label{eqn: extinction_conv}
%\end{equation}

\begin{equation}
    \frac{I_2}{I_1} = \text{OPT}_{\text{Transmission}} = e^{-\frac{A_{\text{atmos}}x}{2.5}}
    \label{eqn: extinction_conv}
\end{equation}

where $I_1$ and $I_2$ are the intensity before and after atmospheric transmission respectively, $A_{\text{atmos}}$ is the atmospheric extinction values from GEMINI, and $x$ is airmass. The computed OPT transmission data and GEMINI NIR transmission data are then applied to the spectrum. For flux to photon conversion Eq. \ref{eqn: photon_conv} is used which assumes flux $F$ in units of erg cm$^{-2}$ s$^{-1}$  $\text{\r{A}}^{-1}$, 

\begin{equation}
    N_{\text{photons}} = \frac{F \  \Delta \lambda \ A_{\text{mirror}} \ t_{\text{exposure}} }{E_{\text{photon}}} 
    \label{eqn: photon_conv}
\end{equation}

where $\Delta \lambda$ is the size of the wavelength bins in the spectrum, $A_{\text{mirror}}$ is the area of the telescope mirror, $t_{\text{exposure}}$ is the exposure time, $N_{\text{photons}}$ is the number of photons arriving within $\Delta \lambda$ with wavelength $\lambda$, and the energy of the photon is defined as $E_{\text{photon}} = hc \text{/} \lambda$. Telescope and optics reflectance are modelled using material reflectivity values as a function of wavelength. For the telescope mirrors and optical surfaces transmission, freshly recoated silver mirror transmission data is used from the GEMINI telescopes\footnote{\url{http://www.gemini.edu/observing/telescopes-and-sites/telescopes}}.

\subsection{Instrument slit and MKID QE}\label{Ins_slit}
This section of KSIM is where the effects of the incoming light propagating through the slit are simulated. A transmissions file is used containing transmission factors with respect to wavelength, which can be generated within KSIM.

\begin{figure}
\includegraphics[width=\columnwidth]{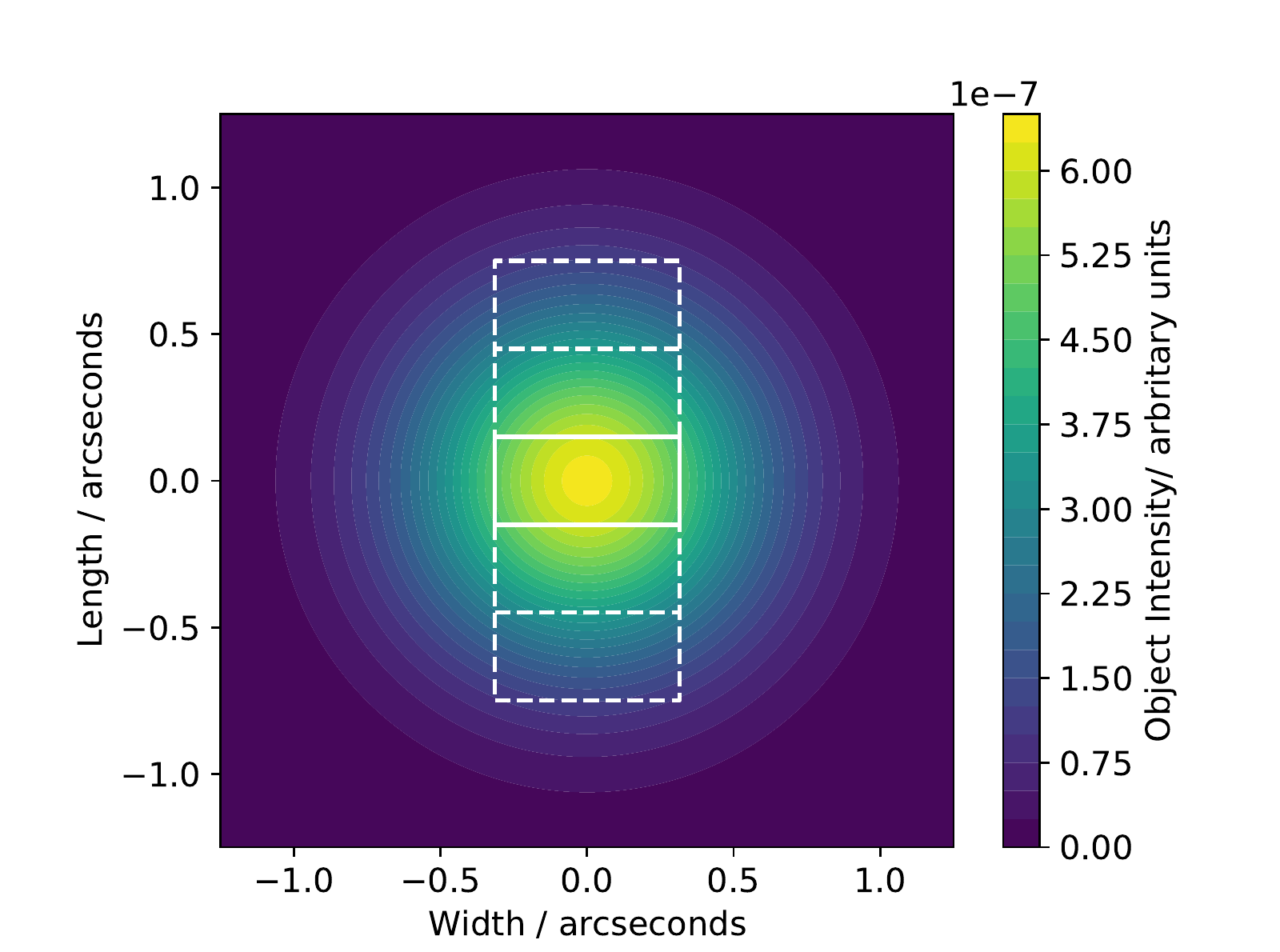}
    \caption{Gaussian distribution of a stellar object because of simulated atmosphere. Parameters were; seeing set to 0.8", airmass 1.5, slit width 0.63" and length 2.7", pixel scale of 0.3", and at a wavelength of 534nm. The solid box shows the spaxel used in this work. The dashed boxes represent a potential spatial geometry using additional spaxels, e.g. for simultaneous sky measurement.}
    \label{fig: gaussian_fields}
\end{figure}

A Gaussian distribution is used to model the point spread function of the incoming point source object flux, over the desired slit parameters. To create one of the transmission files the standard deviation of this Gaussian is calculated. An example of this Gaussian distribution is shown in Fig. \ref{fig: gaussian_fields}. 

The standard deviation is dependant on the wavelength of the incoming photons and is also the result of atmospheric effects, i.e. seeing in units of arcseconds. The simulation calculates the standard deviation value by finding the FWHM using a method outlined in \cite{vandenAncker2016} and is also used in the X-Shooter Exposure Time Calculator\footnote{\url{https://www.eso.org/observing/etc/bin/gen/form?INS.NAME=X-SHOOTER+INS.MODE=spectro}}. To determine the FWHM of the Gaussian, the Image Quality (IQ) using Eq. \ref{eqn: IQ} is found,

\begin{equation}
    \text{IQ} = \sqrt{\text{FWHM}^2_{\text{atm}}(s,X,\lambda) + \text{FWHM}^2_{\text{tel}}(D,\lambda) + \text{FWHM}^2_{\text{ins}}}
    \label{eqn: IQ}
\end{equation}

\noindent where $s$ is seeing in arcseconds, $x$ airmass, $\lambda$ wavelength in nm, and $D$ telescope diameter in metres. The value of $\text{FWHM}_{\text{ins}}$ is a constant adopted from \cite{vandenAncker2016} as an estimate for KIDSpec also. $\text{FWHM}_{\text{tel}}$ is calculated as shown in Eq. \ref{eqn: FWHM_tel} which is the FWHM of the Airy disc, 

\begin{equation}
    \text{FWHM}_{\text{tel}}(D,\lambda) = 0.000212\frac{\lambda}{D}.
    \label{eqn: FWHM_tel}
\end{equation}

\noindent Finally,

\begin{equation}
\begin{split}
\text{FWHM}_{\text{atm}}(s,x,\lambda) = {} & s \times x^{0.6} \times \frac{\lambda}{500nm}^{-0.2} \times \\
    & \sqrt{1 + F_{\text{Kolb}} \times 2.183 \times (r_0/L_0)^{0.356}}
    \label{eqn: FWHM_atm}
\end{split}
\end{equation}

\noindent where $L_0$ is the wavefront outer-scale, $F_{\text{Kolb}}$ is the Kolb factor, and $r_0$ is the Fried parameter. When $L_0$ is approached to or exceeded by the telescope's diameter, the optical turbulence effects change. For example tip and tilt aberration power is reduced \citep{Ziad2016}. The Fried parameter is defined as the telescope diameter required for its diffraction limited resolution to be equal to the seeing resolution \citep{Martinez2010} as described in Eq. \ref{eqn: fried},

\begin{equation}
    r_0 = 0.976 \times 500.0 \times 10^{-9} nm/s \times \frac{180 \times 3600}{\pi} \times \frac{\lambda}{500.0}^{1.2} \times x^{-0.6}.
    \label{eqn: fried}
\end{equation}

\noindent The Kolb factor is defined as follows, 

\begin{equation}
    F_{\text{Kolb}} = \frac{1}{1 + 300 \times D/L_0} - 1
    \label{eqn: kolb}
\end{equation}

\noindent with the same definitions as above. When the standard deviation is calculated and applied to a Gaussian distribution in a 2D plane, it results in images similar to Fig. \ref{fig: gaussian_fields}. This image shows the spread of a point source's incoming flux to ground observatories as a result of atmospheric effects. These Gaussian fields are used to find the object intensity transmission through the user-set slit, depending on wavelength.

At this point in the simulation the sky background is added, using data generated from the ESO SKYCALC Sky Model Calculator\footnote{\url{http://www.eso.org/observing/etc/bin/gen/form?INS.MODE=swspectr+INS.NAME=SKYCALC}}. A separate sky-only simulation is computed simultaneously from this point, for sky subtraction in the final steps of KSIM. This assumes the same slit geometry as the science observation. In practice, this could be taken by offsetting the telescope at the expense of additional telescope time, or a simultaneous sky exposure could be taken using additional MKIDs as described in Sec. \ref{kidspec_design}.

The QE of the MKIDs is also applied here to simplify the software at later stages. The QE value used is either a constant value of 0.75 \citep{Marsden2013}, or a value between 0.73 at 200nm reducing to 0.22 at 3000nm, as shown in \cite{Mazin2010}.

\subsection{Grating}\label{Grating}
KIDSpec will use an echelle grating for multiple orders to be exposed onto the MKID array. The grating of the instrument and its efficiency is simulated using a method outlined in \cite{Casini2014}.

This section of the simulation is split into three parts; calculating the relevant orders and efficiency of the grating setup, assigning the pixels that the spectrum photons fall onto by their wavelength, and distributing the photons which arrive to the instrument. For the first part the simulation calculates each order's wavelengths up to a chosen order number. While the light from all orders will land onto the MKIDs, only orders with wavelengths within the bandpass are considered. The wavelengths KIDSpec will observe in KSIM are calculated using the grating equation,

\begin{equation}
    \lambda = \frac{d\left(\sin(\beta) + \sin(\alpha)\right)}{m}
    \label{eqn: grating_eq}
\end{equation}

\noindent where $\lambda$ is wavelength, $d$ the groove spacing, $m$ the order number,  $\alpha$ the incidence angle, and $\beta$ the angle of diffraction. The orders which have values of $\lambda$ within the bandpass are extracted.

The efficiency of the grating is found using Eq. \ref{eqn: casini},

\begin{equation}
    I(\beta) = \text{sinc}^2\left(m\pi\frac{\cos(\alpha)}{\cos(\alpha - \phi)}\frac{\sin(\alpha - \phi) + \sin(\beta - \phi)}{\sin(\alpha) + \sin(\beta)}\right)
    \label{eqn: casini}
\end{equation}

\noindent where $\phi$ is the blaze angle and all other variables have the same definition as Eq. \ref{eqn: grating_eq}. The calculated wavelengths (from Eq. \ref{eqn: grating_eq}) and efficiencies are mapped to the MKIDs, depending on their position in the array. The number of wavelengths a spectral pixel (an MKID) will observe is the number of orders passing the grating calculations. For the third part of this section, the majority of input data spectra used will have a different spectral resolution than what KSIM can produce. This is because of a large number of effective MKID pixels; the data spectrum can be interpolated if necessary. Using this spectrum, each of the spectrum's wavelengths are binned onto a KIDSpec grid of effective MKID pixel bins. Each wavelength bin of photons has the grating efficiency of the pixel bin it is placed in, applied to it during this process. 

A Poisson distribution is used to simulate the photon shot noise during an observation. We are also able to model dead MKID pixels as a result of current fabrication methods \citep{Walter2020}. These dead pixels are unable to detect incoming photons, and so dead pixels have their photon counts set to zero here. The proportion of dead pixels can be set by the user.

\subsection{MKID Order Gaussians}\label{pix_gaussian}
The MKID pixel response to each order incident on it is applied in this section of the simulation tool, examples of which are shown in Fig. \ref{fig: order_resolving_energy_res}. Each MKID has a single wavelength incident on it from each order, defined here as the order wavelength. However incoming photons of the same wavelength cause a range of responses by the MKID, resulting in a range of measured wavelengths for each order, as discussed in Sec. \ref{kidspec_design}. The size of this range is dependant on the $R_{E}$. To represent this range of wavelengths, each order observed by the MKID is assigned a random Gaussian distribution. The mean of this order Gaussian is the order wavelength incident on that MKID. The standard deviation value of the order Gaussian distribution is calculated using Eq. \ref{eqn: 1sig}, used from \cite{Marsden2013} with the FWHM term defined as in Sec. \ref{kidspec_design},

\begin{equation}
    1\sigma = \frac{\mu_{\lambda}}{R_E \times 2\sqrt{2\log(2)}}
    \label{eqn: 1sig}
\end{equation}

with $R_{E}$ and the mean wavelength $\mu_{\lambda}$. $R_{E}$ in the MKID array reduces as the wavelength increases, owing to them breaking fewer Cooper pairs, which also causes the fast rise heights to drop closer to the phase stream noise. In the simulation the $R_{E}$ drops from the value set in the parameters, decreasing linearly with increasing wavelength \citep{OBrien2014}. $R_{var}$ is implemented here if desired, and will vary each MKID pixel's $R_E$ following an approximately Gaussian trend \citep{Meeker2018} centred on the set $R_E$ in the KSIM parameters. The scale of $R_{var}$ can be set by the user.

Order overlap within the simulation is defined as when the Gaussians calculated above, overlap in wavelength as seen in Fig. \ref{fig: order_resolving_energy_res}. The order of a photon within an overlapping region of wavelength, is determined by the order with the highest probability for that particular wavelength. After the overlap is calculated, the photons are sorted into their highest probability orders using the description above. A list of photon counts seen by the MKID from each order is then produced. A non-trivial scenario which can be seen with KSIM is when the Gaussians are not all equal height, which can negatively impact the percentage of photons misidentified. An example of this is bright sky lines which could cause this difference in heights to occur during observations. 

For the bottom panel of Fig. \ref{fig: order_resolving_energy_res} where the $R_{E}$ was 30 at 400nm, the misidentified photon percentage was 0.2, 0.2, and 0.1 \% for orders 8, 9, and 10 respectively. When the incoming photon count of order 9 was increased to be 100 times the counts of the other orders, orders 8, 9, and 10 had misidentified photon percentages of 5.8, 0.1, and 7.1\%  – a decrease in misidentified photons for the more intense order 9, and an increase for the adjacent two orders.

The impact of the unequal Gaussian heights can be reduced by careful design of the MKID arrays and optics used for KIDSpec. For example, in a constructed KIDSpec there will be MKID pixels which will fall outside of multiple spectral order's free spectral ranges (FSRs), and as such will receive less flux from these orders. This will occur for MKIDs on the ends of the linear array. If these MKID pixels with fewer orders are where the bright sky lines are exposed then the sky lines will impact fewer orders than if it was exposed onto a central MKID which had wavelengths from all order's FSR.

\subsection{Photon Time Stream}\label{PTS}
An alternative, more computationally expensive method to Sec. \ref{pix_gaussian}, is to simulate the photon time stream (PTS) which each MKID would observe in an exposure. This represents an actual MKID observation more closely than Sec. \ref{pix_gaussian}. This method provides more output information of an object simulation, such as number of potentially saturated time bins. A saturated time bin has had two photons arrive in the same microsecond time bin.

The PTS method produces time bins from an exposure, rather than using only the total photon count in wavelength bins, as described in Sec. \ref{pix_gaussian}. The number of time bins is determined by the exposure time, time resolution of the MKIDs, and any coincidence rejection time the user wishes to incorporate. This coincidence rejection time is incorporated because of the `cool-off' tail of a fast rise in a phase time stream, owing to a photon arrival. This tail can be seen in Fig. \ref{fig: pulse_example}. If a photon then arrives during this tail, the height of the rise will have the tail's height included within it, which makes it non-trivial to find the incoming photon's true fast rise height. Because of this, a duration of time bins following a photon detection are rejected in software. For an example of a time bin calculation, an MKID with time resolution $1\mu s$, used for an exposure of 50s, with a coincidence rejection time of $200\mu s$, would result in a minimum of 250,000 time bins. This $200\mu s$ rejection time is a conservative figure, as analysis of MKID data improves this can be reduced to effectively zero, and currently on MEC a $10\mu s$ rejection time is used \citep{Walter2020}. The photons arriving at each MKID are then assigned randomly into the time stream bins. Each order the MKID observes receives its own time stream. The total MKID response for each time bin is determined by summing the energies of the photons which arrived in that time bin.

%\begin{table}
%\centering
%\caption{Example time bins from a simulation of the photon time stream. Parameters include a 50s exposure onto an MKID with $1\mu s$ time resolution with a $200\mu s$ coincidence rejection time after detection events. The coincidence rejection time is included in the Time column, with the next time step shown after a photon event is after the coincidence rejection time. The coincidence rejection time chosen is a conservative estimate from the top left panel of Fig. \ref{fig: order_resolving_energy_res}.}
%\label{table: photon_time_stream_example_bins}
%\renewcommand\arraystretch{1.5}
%\begin{tabular}{@{} c   c  c  c  @{}}
%\toprule
%Time & Order 1 & Order 2  & Total \\
%($\mu s$) & photon (nm) & photon (nm) & response (nm) 
%\\ \midrule
%1 & 800 & 0 & 800 \\ 
%202 &  0   & 0  & 0 \\
%203 & 0 & 400 & 400 \\ 
%404 & 800$\times$2 & 0 & 400 \\
%605 & 800 & 400 & 267 \\ \bottomrule
%\end{tabular}
%\end{table}

$R_{E}$ of the MKIDs at each order is calculated in the same way as in Sec. \ref{pix_gaussian}, and similarly used in a random Gaussian distribution with the order wavelength as the mean. After the PTS has been generated, the total response of each time bin is taken and KSIM attempts to assign the correct order wavelength to the result of that time bin. A Gaussian probability density function is created for each order, using the order wavelength and previously calculated $1\sigma$ as the mean and standard deviation respectively. The order distribution which produces the highest probability is assumed to be the order of the photon which arrived in that time bin.

A maximum of one photon per time bin is allowed here; any additional photons saturate the time bin and have the potential to cause misidentifications in the assignments of wavelengths to these photons. In the PTS the time bins of the following $\approx100\mu $s after a photon event are not used because of the `cool-off' tail following the fast rise event. It should be noted that the method used here aims to be conservative in its simulation.

As $R_{E}$ increases with MKID development, more orders will also be able to be resolved, lowering the required MKIDs in an array as discussed in Sec. \ref{kidspec_design}. The potential misidentification issues will also be reduced in the future as the readout technology continues to be developed, and is not an issue with the MKIDs themselves.

\subsection{Simulation Output}\label{sim_out}
Here a grid of the MKID's responses to the simulated observation is produced. This response grid is in dimensions of order number vs pixel, and has a wavelength counterpart. Sky subtraction is applied here using the sky only response grid, which has been run through the simulation tool simultaneously. Using the sky subtracted and sky only grids the SNR is determined.

Orders are merged using these response grids by rebinning the response grids onto a larger grid to form a 1D spectrum. From the rebinning a raw photon spectrum is produced, before having a set of standard star weights applied to the wavelength bins. The observed photon counts are not discarded however and can be retrieved. These standard star weights are calculated by running a standard star, GD71, through the simulation tool. The output of the standard star simulation is divided by the original standard star spectrum, to produce weights with respect to wavelength to account for the instrument, atmosphere and telescope response, similarly to \cite{Modigliani2010}. Instead of using the known values of the various responses in KSIM, the standard star weights are used to represent an observation using KIDSpec where these responses may not be known. The weighted spectrum is then converted back to units of flux in $erg \, cm^{-2} \, s^{-1} \, \text{\textit{\AA}}^{-1}$.

%--------------------------------------------------------------------
\section{Simulations}\label{sims}

In this section we optimise potential designs of a medium resolution KIDSpec and show the impacts MKID fabrication effects may have on observations. Consistent simulation parameters are shown in Table \ref{table: parameters_runs}.

\subsection{KSIM to influence KIDSpec's design}\label{kidspec_adapt}

\begin{table}
\centering
\caption{Consistent parameters for simulations shown in this paper. These parameters were used for each of the three setups described in Sec. \ref{kidspec_adapt}, which were used to simulate the cases in this paper, unless stated otherwise. The QE used was from \protect\cite{Mazin2010}.}
\label{table: parameters_runs}
\renewcommand\arraystretch{1.5}
\begin{tabular}{@{} c    c @{}}
\toprule
Parameter & Value\\ \midrule
Seeing (arcsec) & 0.8 \\
Airmass & 1.5 \\
Slit width (arcsec) & 0.63 \\
Pixel FOV (arcsec) & 0.3 \\
Incidence angle (deg) & 25 \\
Blaze angle (deg) & 25 \\ 
Telescope diameter (m) & 8 \\
Effective area ($\text{m}^2$) & 50 \\
QE & 0.73-0.22 \\
\bottomrule
\end{tabular}
\end{table}

\begin{figure*}
\centering
\includegraphics[width=\columnwidth]{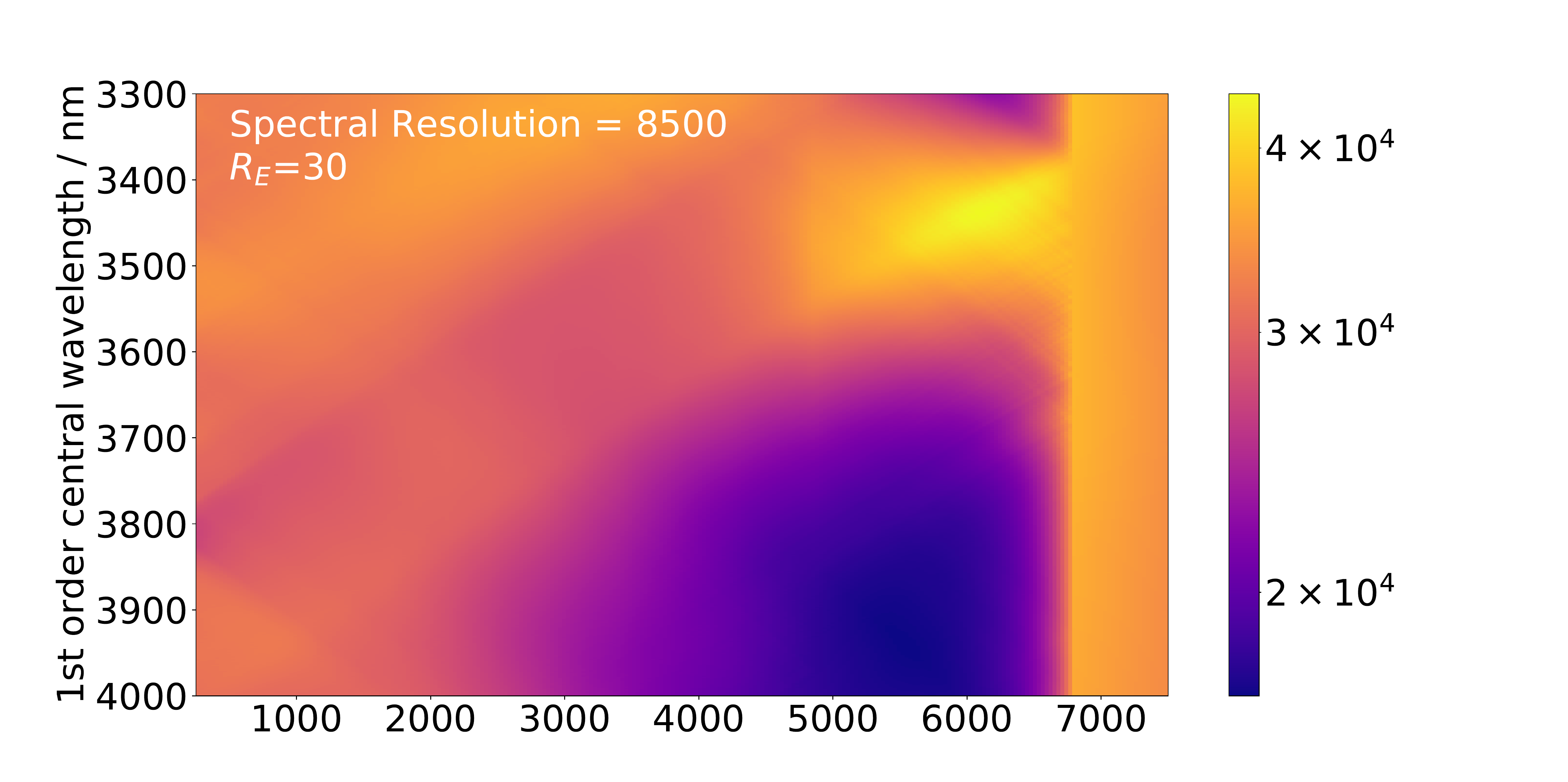}
\includegraphics[width=\columnwidth]{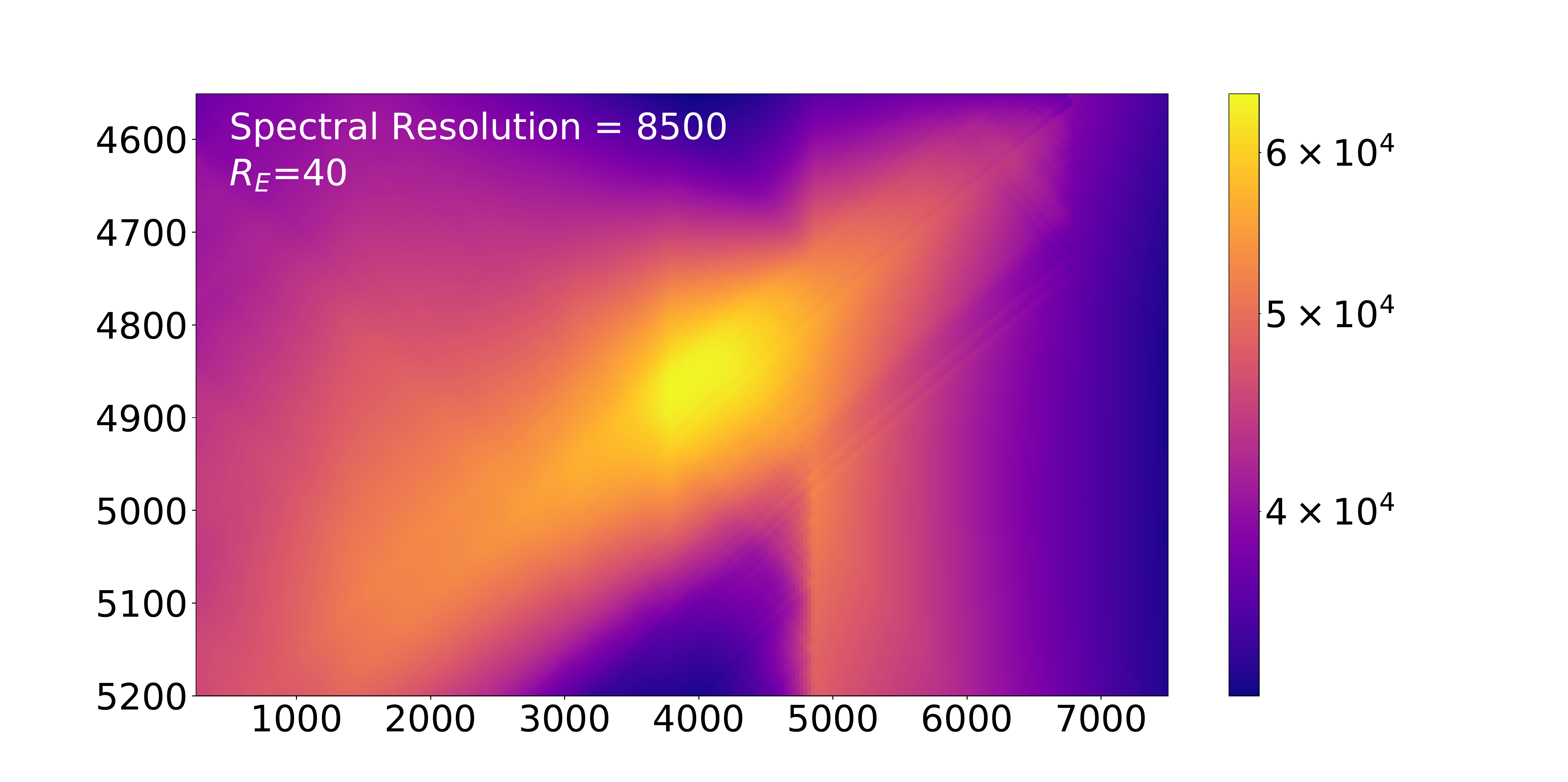}
\includegraphics[width=\columnwidth]{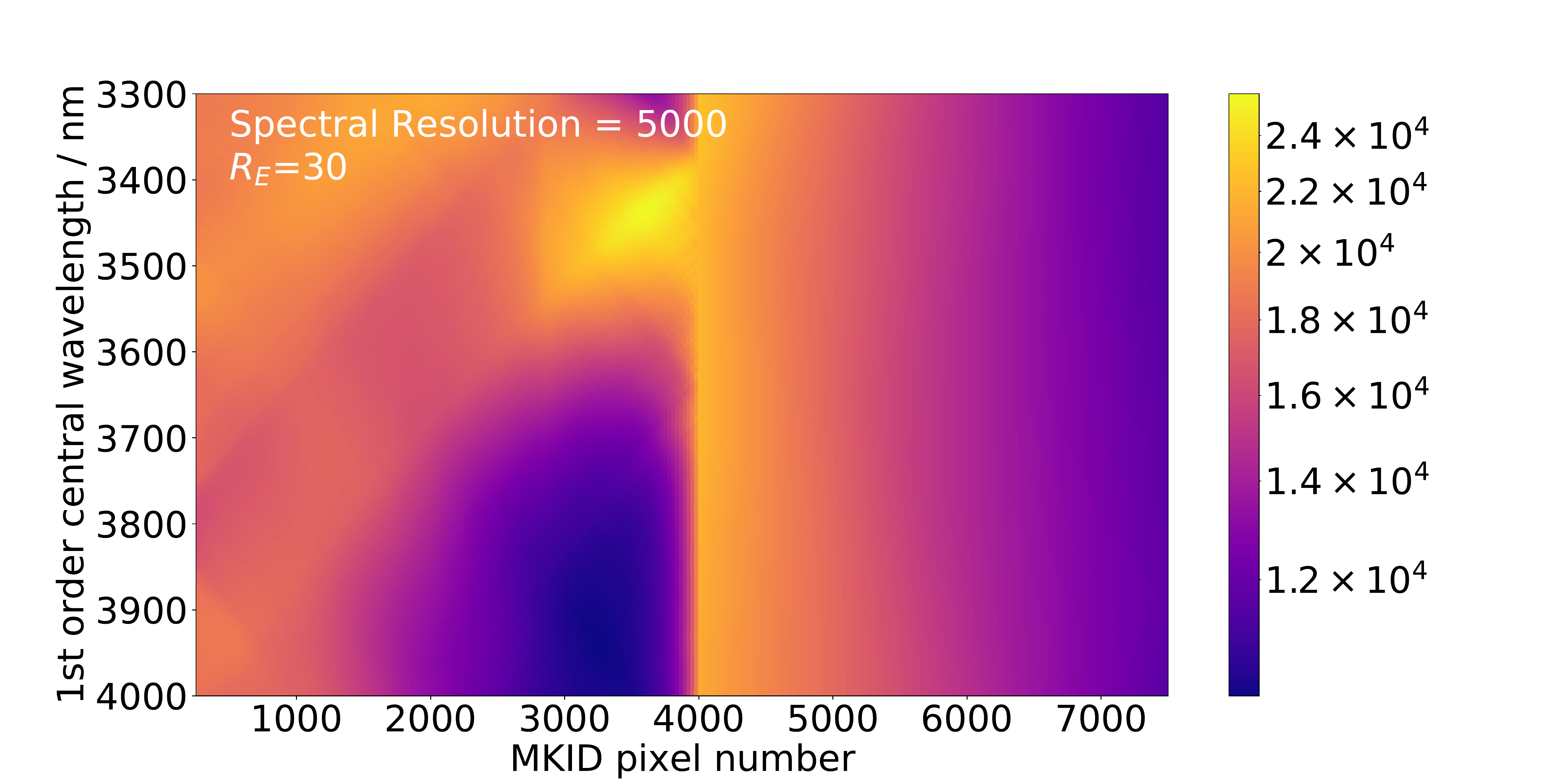}
\includegraphics[width=\columnwidth]{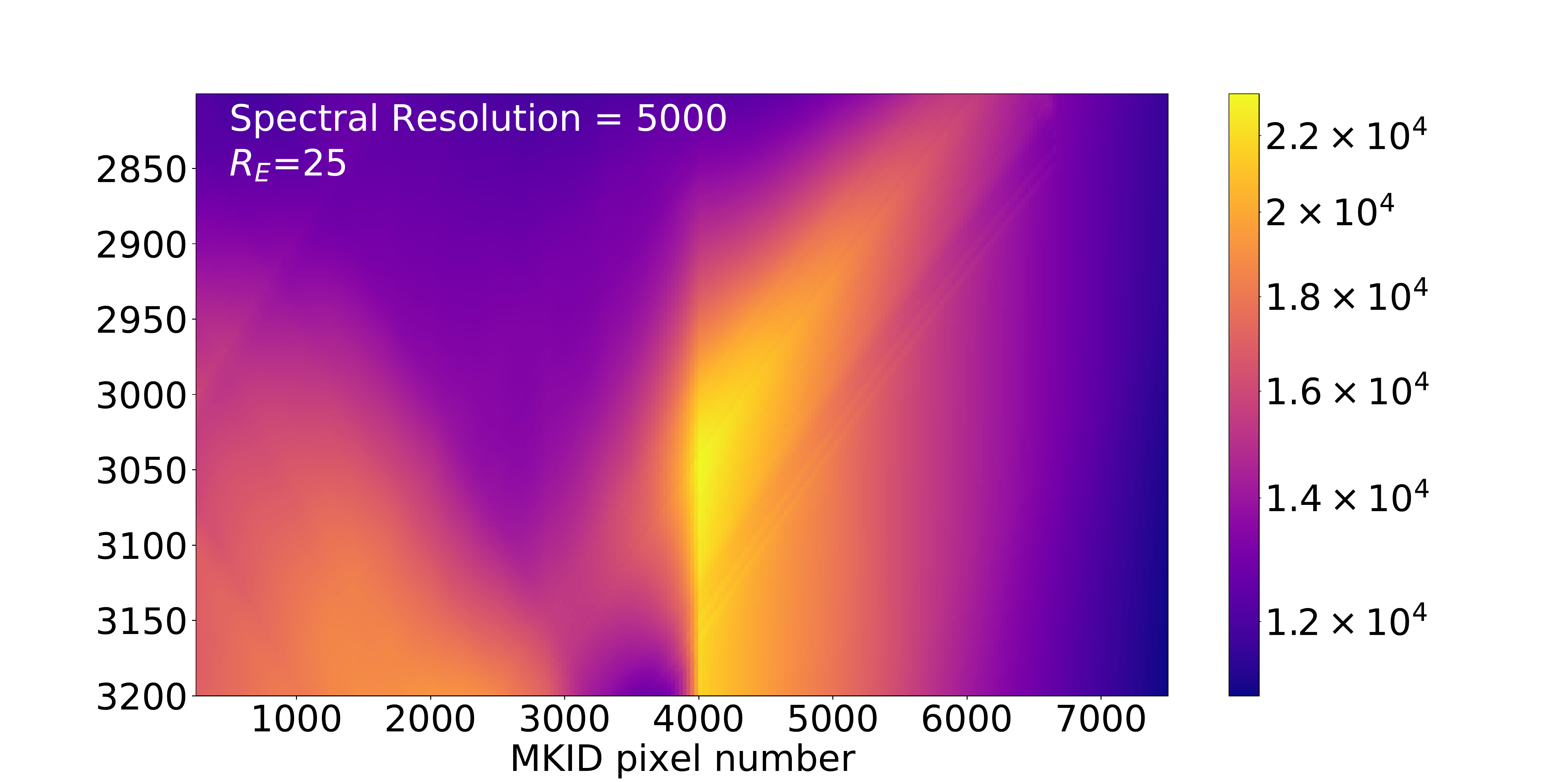}
    \caption{ Grating order placement optimiser maps for a KIDSpec with spectral resolutions of 8500 at $R_{E}$s of 30 (\textbf{Top Left}) and 40 (\textbf{Top Right}) and 5000 at $R_{E}$s of 30 (\textbf{Bottom Left}) and 25 (\textbf{Bottom Right}). Sampled were the first order central wavelength and number of MKIDs required. Score, indicated by colour in these plots, was gained by having wavelength coverage in areas of low sky brightness and high atmospheric transmission, while maximising coverage. Common to all plots is a vertical line which signifies the point where adding MKIDs no longer improves the bandpass coverage and there is unnecessary bandpass overlap in the orders. The large area of low score for the (\textbf{Top Left}) plot in the area of $\approx{5000}$ MKIDs is the result of the positions of the orders being in poor areas of the bandpass and experiencing more overlap. Similarly for (\textbf{Bottom Left}) plot which has the same $R_{E}$.}
    \label{fig: optimiser_plot}
\end{figure*}

\begin{figure*}
\centering
\includegraphics[width=\columnwidth]{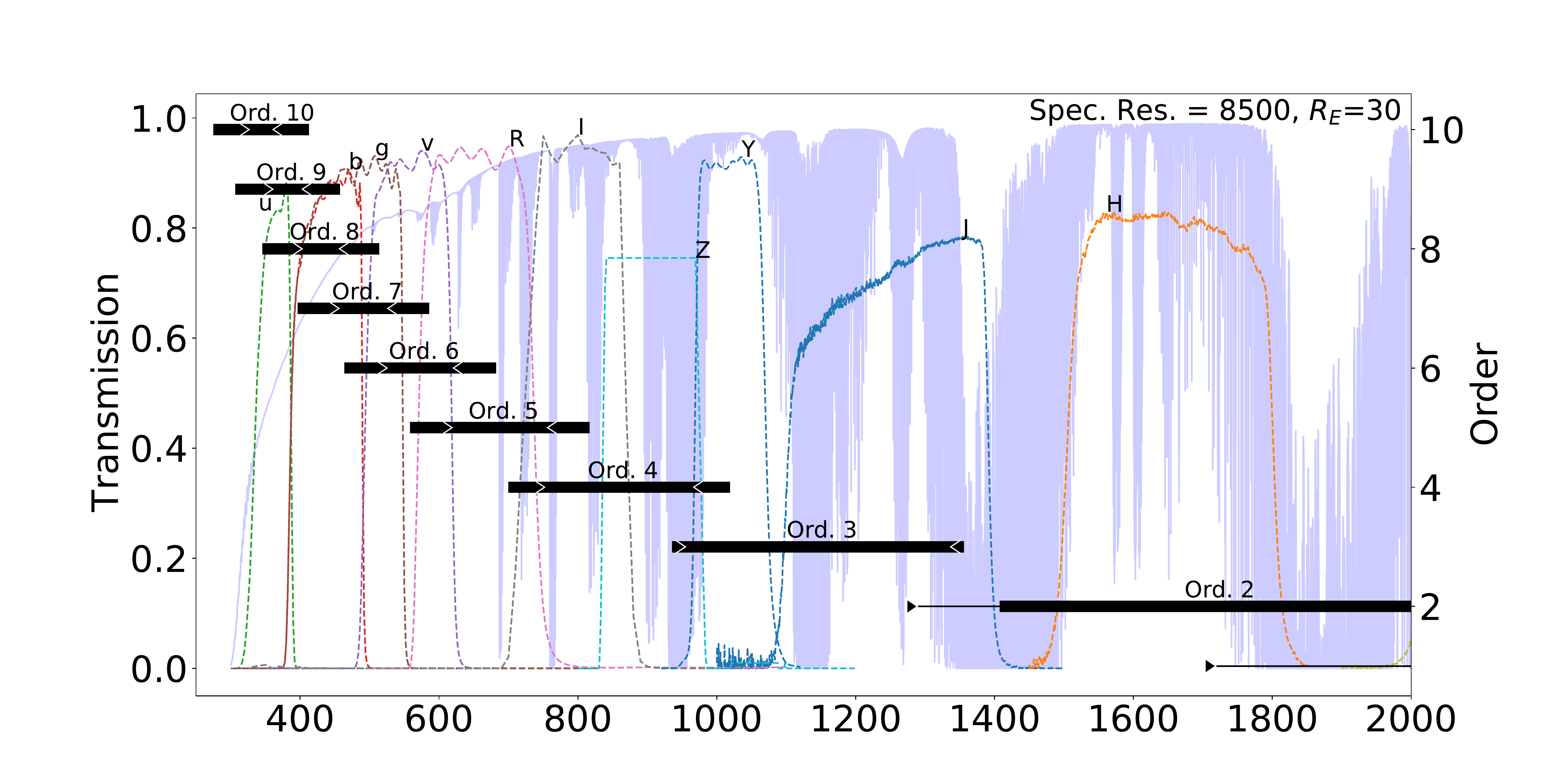}
\includegraphics[width=\columnwidth]{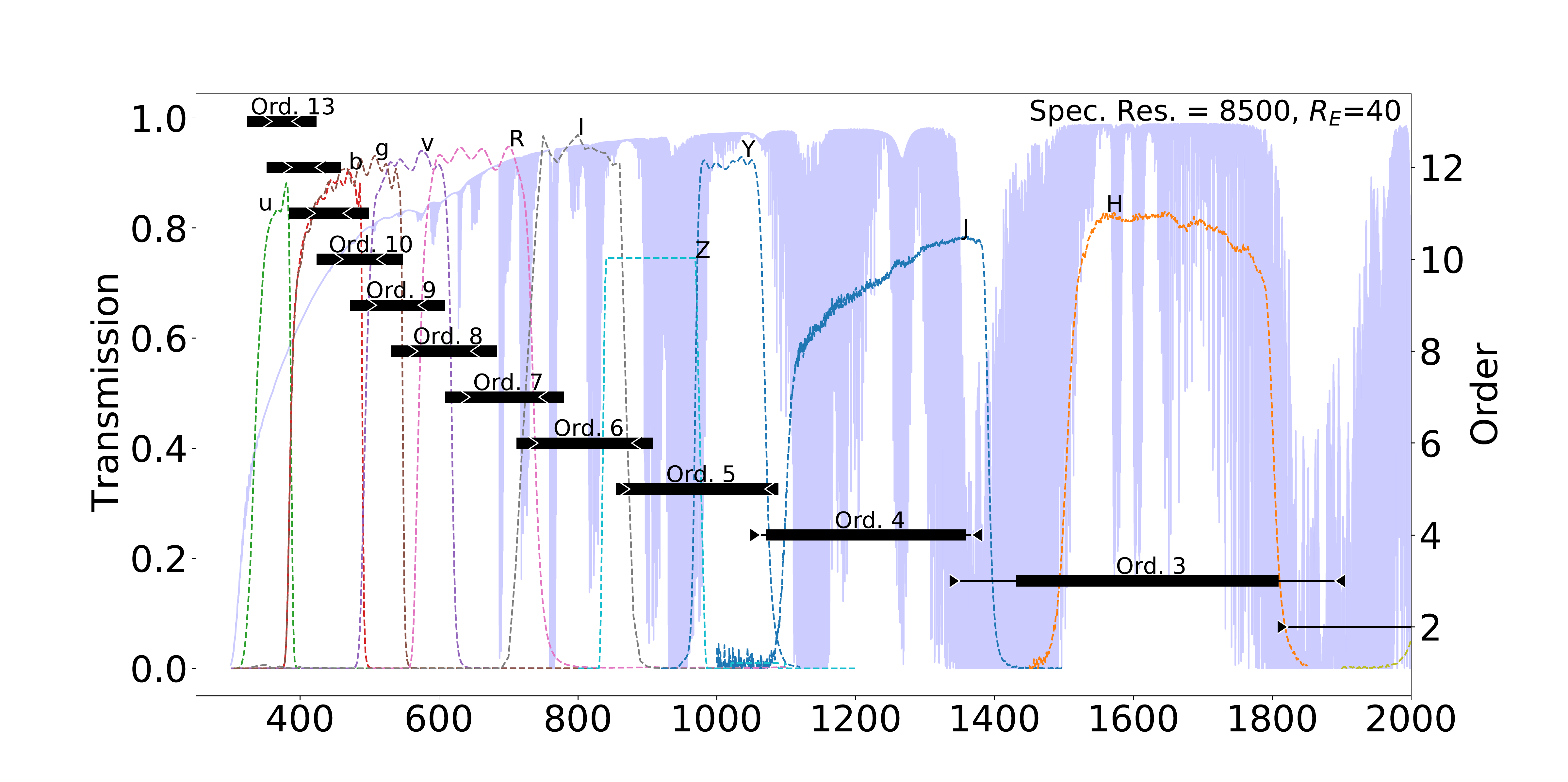}
\includegraphics[width=\columnwidth]{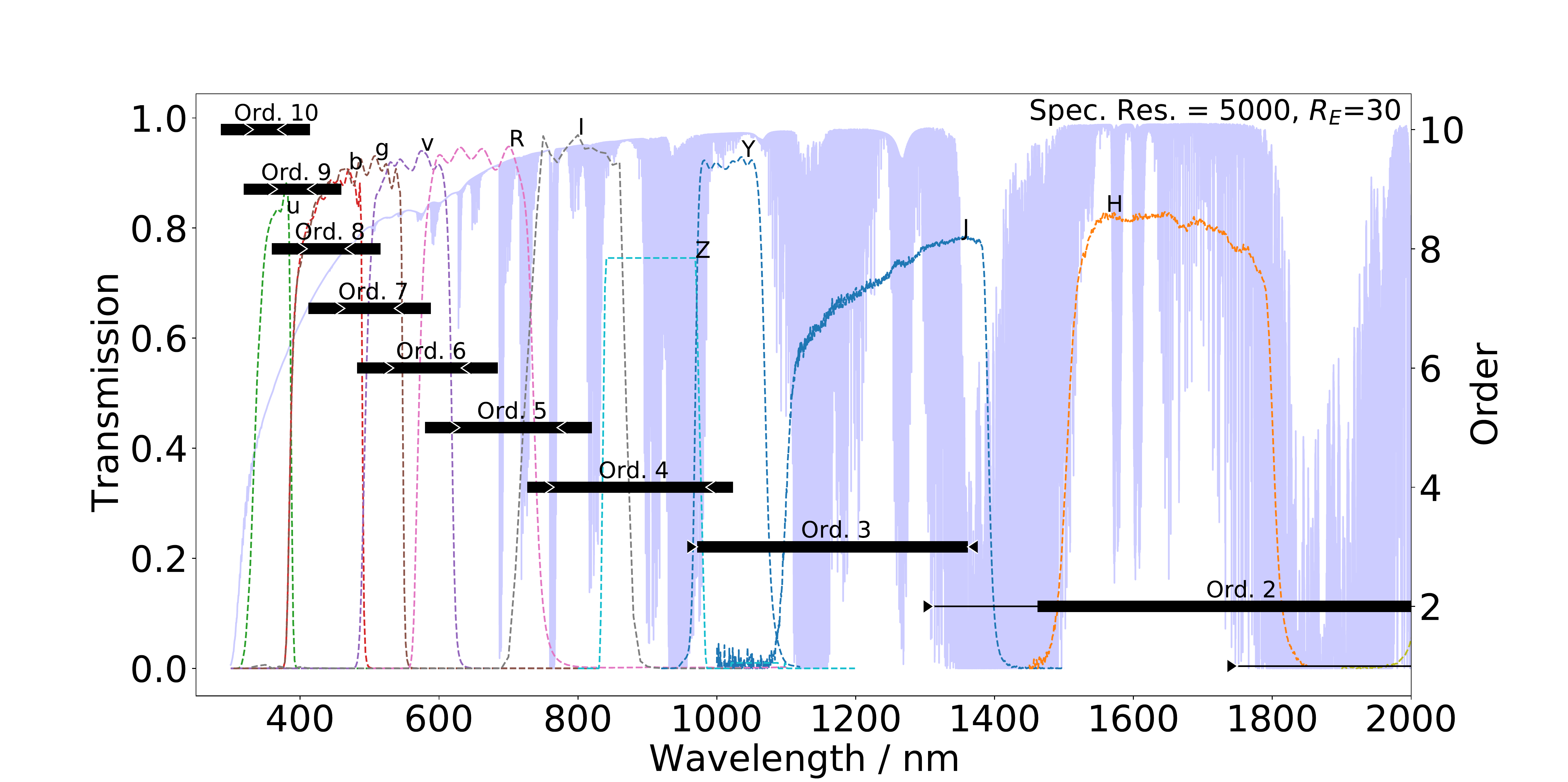}
\includegraphics[width=\columnwidth]{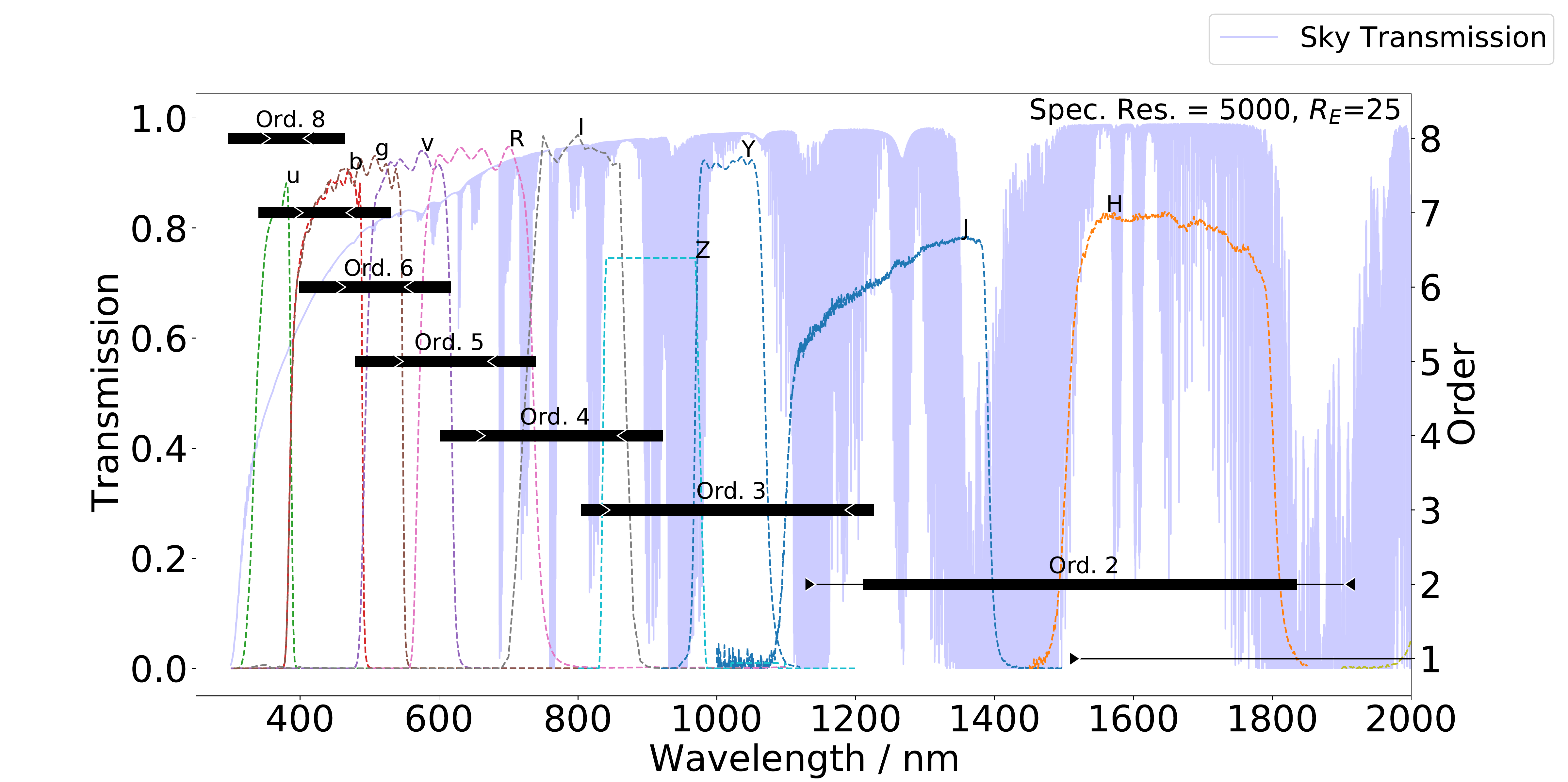}
    \caption{ Grating order placement optimiser results for a KIDSpec with spectral resolutions of 8500 at $R_{E}$s of 30 (\textbf{Top Left}) and 40 (\textbf{Top Right}) and 5000 at $R_{E}$s of 30 (\textbf{Bottom Left}) and 25 (\textbf{Bottom Right}). Plotted for all are the grating order wavelength ranges observed by the MKIDs in bold black bars. The free spectral range for each order is represented by the thinner black lines and arrows pointing inwards. Magnitude bands from ESO used for ETC simulations are also plotted, with the GEMINI atmospheric transmission data.}
    \label{fig: scorer_res}
\end{figure*}
%\begin{figure*}
%\centering
%\includegraphics[width=\columnwidth]{Images2/lim_mags_8500_re30.pdf}
%\includegraphics[width=\columnwidth]{Images2/lim_mags_8500_re40.pdf}
%\includegraphics[width=\columnwidth]{Images2/lim_mags_5000_re30.pdf}
%\includegraphics[width=\columnwidth]{Images2/lim_mags_5000_re25.pdf}
%    \caption{Predicted limiting magnitude results for a single-arm KIDSpec with spectral resolutions of 8500 at $R_{E}$s of 30 (\textbf{Top Left}) and 40 (\textbf{Top Right}) and 5000 at $R_{E}$s of 30 (\textbf{Bottom Left}) and 25 (\textbf{Bottom Right}). The FORS and the KIDSpec diamond marker limiting magnitudes are for SNR>5, with all others being for SNR>10. For all areas but the $\approx{400}$nm section in the 8500 spectral resolution result, these setups improve on X-Shooter's limiting magnitude. The 5000 setup for an hour exposure with SNR>5 even approaches the limiting magnitude for the FORS instrument of the same threshold. }
%    \label{fig: lim_mags_1arm}
%\end{figure*}

\begin{table*}
\centering
\caption{Predicted limiting magnitude results for a KIDSpec with spectral resolutions of 8500 at $R_{E}$s of 30 and 40, and 5000 at $R_{E}$s of 30 and 25. The wavelengths shown are the blaze wavelengths for each spectral order. An exposure of 3600s and 30s on an 8m diameter telescope were used for these simulations with a SNR>10 threshold. In the AB magnitude columns the exposure times are separated in the form of 3600s/30s.}
\label{table: lim_mags}
\renewcommand\arraystretch{1.5}
\begin{tabular}{@{}cccc ccc cccc@{}}
\toprule
\multicolumn{4}{c}{Spectral Res. 8500} &&&& \multicolumn{4}{c}{Spectral Res. 5000} \\
\multicolumn{2}{c}{Energy Res. 40} & \multicolumn{2}{c}{Energy Res. 30} &&&& \multicolumn{2}{c}{Energy Res. 30} & \multicolumn{2}{c}{Energy Res. 25} \\
Wavelength/nm&AB mag.&Wavelength/nm&AB mag. &&&& Wavelength/nm&AB mag.&Wavelength/nm&AB mag.\\\midrule

1617 & 19.0/15.8	&	1742  & 19.5/15.9	&&&&	1741 & 19.8/16.4 &		1507 &	19.0/15.9 \\
1212	& 20.2/16.5	&	1161  & 20.0/16.2	&&&&	1160	& 19.6/15.8 &		1005 &	20.9/17.5 \\
970	& 21.0/17.3	&	871   & 21.3/17.6	&&&&	870	& 21.8/18.2 &		753  &	21.6/17.9 \\
808	& 21.5/17.7	&	697	 & 21.4/17.5	&&&&	696	& 21.7/18.0 & 	603  &	21.4/17.6 \\
693	& 21.7/17.7	&	581	 & 21.0/17.1   &&&&	580 & 21.3/17.5 &		502  &	21.6/17.6 \\
606	& 21.1/17.2	&	498	 & 20.8/16.5	&&&&	497	& 21.4/17.3 & 	431  &	20.8/16.3 \\
539	& 20.8/16.6	&	435	 & 20.3/15.8	&&&&	435 & 20.6/16.3 &		377  &	17.7/12.8 \\
485	& 21.0/16.8	&	387	 & 18.1/13.1	&&&&	387 & 18.6/13.7 &				& \\
441	& 20.2/15.8	&			 &					&&&&			&                 &             & \\
404	& 18.9/14.1	&			&					&&&&		    &                 &             & \\
373	& 16.8/11.6	&			&					&&&&	        &                 &             & \\\bottomrule
\end{tabular}
\end{table*}

When designing KIDSpec the wavelength coverage must be carefully chosen to maximise the use of the MKIDs, covering as wide a desired bandpass as possible. To find the optimal use of a KIDSpec design, a scoring script was used which determined the optimal grating order placements, considering the sky background and atmospheric transmission. The orders were placed using a method from \cite{Cropper2003}. The scorer varied the position of the highest order and number of MKID pixels required given a set spectral resolution and $R_{E}$. Given a certain spectral resolution, the wavelengths in each order were then sampled in the scorer with score gained for the wavelength residing in an area of high atmospheric transmission and low sky background. The atmospheric transmission can be seen in Fig. \ref{fig: scorer_res}. For the atmospheric transmission score, values were given by the transmission values themselves, with score also gained from orders missing areas of low transmission. For the sky background the values were normalised, and then the inverse was used to calculate the score. These results were then summed. Two spectral resolutions were chosen here, 5000 and 8500 to coincide with X-Shooter spectral resolutions \citep{Vernet2011}. A range of $R_{E}$ between 25 and 40 inclusive at approximately 400nm were also selected.

The optimiser's results for the parameter space which was sampled is shown in Fig \ref{fig: optimiser_plot}. For the 8500 spectral resolution setup, two $R_{E}$s were used, 30 and 40. These optimisations resulted in a 1st order central wavelength of 3.44 and 4.86$\mu m$ respectively. The number of pixels determined were 6000 and 3800 respectively. The difference in number of pixels between the two were the result of the $R_{E}=40$ setup observing more grating orders with a $3\sigma$ separation as discussed in Sec. \ref{design}. The gaps in coverage for these setups were optimised by the scorer to contain the large dips in atmospheric transmission, rather than add extra MKIDs to get full bandpass coverage at this resolution. For the 5000 spectral resolution setup $R_{E}$s of 25 and 30 were used. This gave a 1st order central wavelength of 3.04 and 3.50 $\mu m$ respectively, with 4000 and 3200 MKID pixels. Shown in Fig. \ref{fig: scorer_res} are the scorer's chosen grating order placements for a spectral resolution of both 8500 and 5000 at various $R_{E}$s. Magnitude bands from ESO\footnote{\url{https://www.eso.org/observing/etc/doc/formulabook/node12.html}} used for ETC simulations are also plotted in Fig. \ref{fig: scorer_res}, with the GEMINI atmospheric transmission data. The $R_{E}=30$ results demonstrate a trade off between spectral resolution and number of MKID pixels, to achieve similar grating order placement. This trade off is shown by the 8500 spectral resolution results for $R_E =30$ having a number of MKID pixels that was higher, than that for the 5000 spectral resolution. Additionally from Fig. \ref{fig: scorer_res} with the results using the same spectral resolution, a higher $R_E$ allows for a reduction in the number of MKID pixels required for an optimised coverage in the range $\approx{400}-2000$nm. This is understandable, owing to a higher $R_E$ meaning more orders can be separated and so more order wavelengths can be observed on a single MKID.

For all of these KIDSpec setups, limiting magnitudes were predicted using KSIM. The AB magnitudes of each order's blaze wavelength were simulated, for varying exposures on a 8m class telescope, at the point where SNR>10, given a 30s exposure. This demonstrates KIDSpec's potential for short exposure science. A one hour exposure for SNR>10 was also done, to match the X-Shooter limiting magnitudes setup in \cite{Vernet2011}. Table \ref{table: lim_mags} contains the results of the predictions. From Table \ref{table: lim_mags} KIDSpec could, for exposures of only 30 seconds for an SNR>10, achieve magnitudes of up to $\approx{17.7}$ for a spectral resolution of 8500, and $\approx{18.0}$ for 5000. This increases to up to $\approx{21.7}$ and $\approx{21.8}$ for the two spectral resolutions respectively, for an exposure of one hour.

KIDSpec's spectral resolution can be lowered post observation with rebinning and no detriment, owing to the lack of readout noise. The 5000 spectral resolution performs similarly to the 8500 spectral resolution setup towards the H band. This is the result of the low atmospheric transmission areas contained inside these orders as seen in Fig. \ref{fig: scorer_res}. For the two spectral resolution setups with $R_{E}=30$, as expected, the lower spectral resolution had fainter magnitudes, except for the $\approx{1200}$nm result where the two are very similar. This is because of this wavelength being in a low atmosphere transmission region in the J band. These results demonstrate that KIDSpec has the potential to make a large impact, while still having the flexibility in post observation rebinning and low-noise capabilities of the MKID devices. 

These varying simulated KIDSpec setups demonstrate KSIM's potential to simulate a variety of design choices for KIDSpec and will therefore help finalise a KIDSpec design. 

In the rest of this work, we utilise the two 5000 spectral resolution setups, and the 8500 spectral resolution setup with $R_{E}=40$. The $R_{E}=30$ setup is omitted, because of it having similar grating order placements as its 5000 spectral resolution counterpart.

\begin{figure*}
\includegraphics[width=18cm]{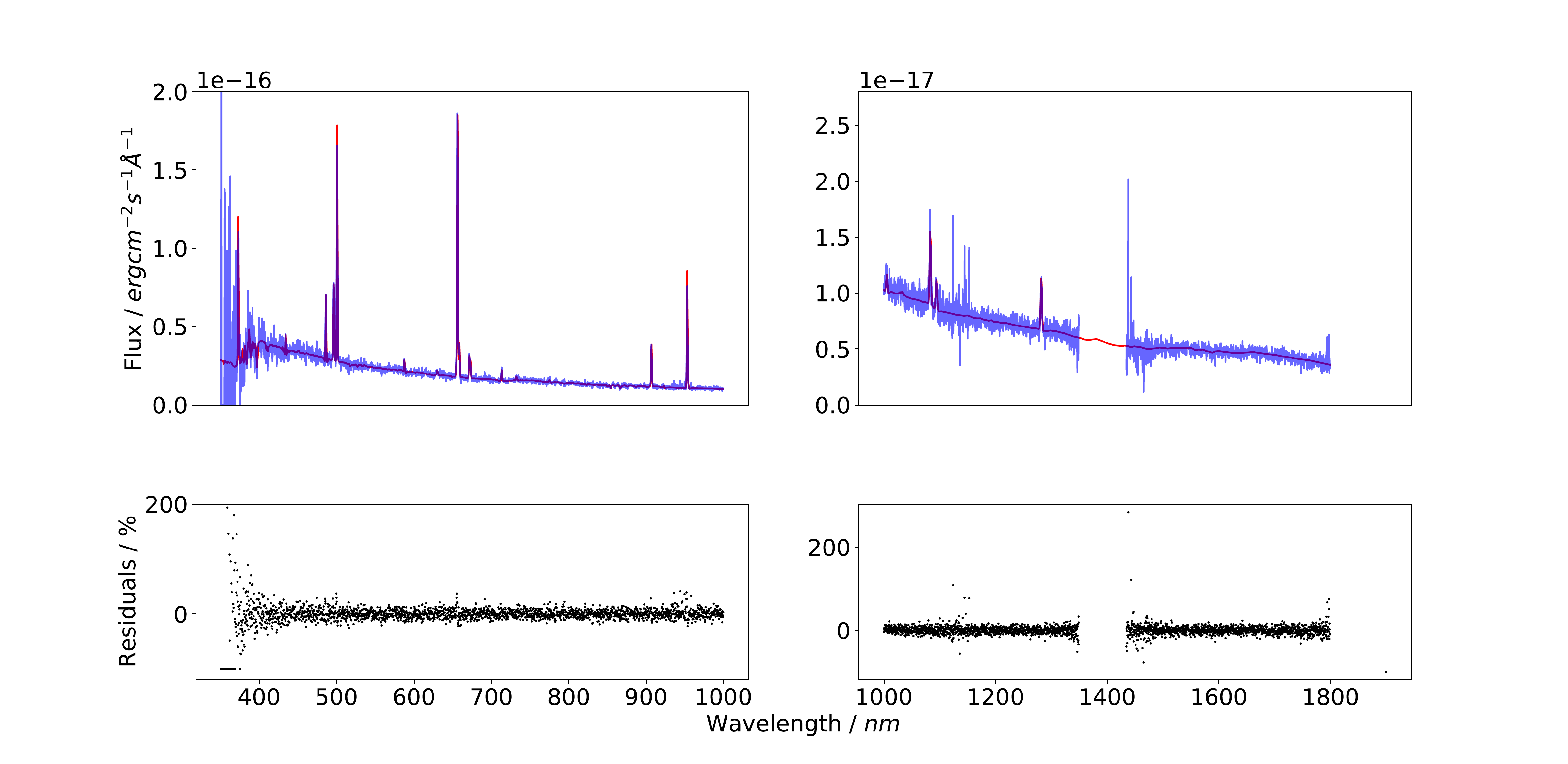}
    \caption{Simulations of a JAGUAR mock spectrum ($m_{R}=20$) for 180s on a 8m telescope using the consistent parameters in Table \ref{table: parameters_runs}. Shown is the result for the KIDSpec setup with a spectral resolution of 8500, rebinned to a spectral resolution of $\approx{4000}$, the original resolution of the JAGUAR spectra. In the upper segment of the Figure the blue represents the KSIM result, and the red the input spectrum. The percentage residuals for their respective simulations are included in the bottom segment of the Figure. The result from KSIM has been split into two plots because of} the lower flux at wavelengths higher than 1000nm, this was simulated for a KIDSpec design with bandpass $0.35-1.8\mu m$.
    \label{fig: jaguar_spec}
\end{figure*}
\begin{figure*} 
\includegraphics[width=\columnwidth]{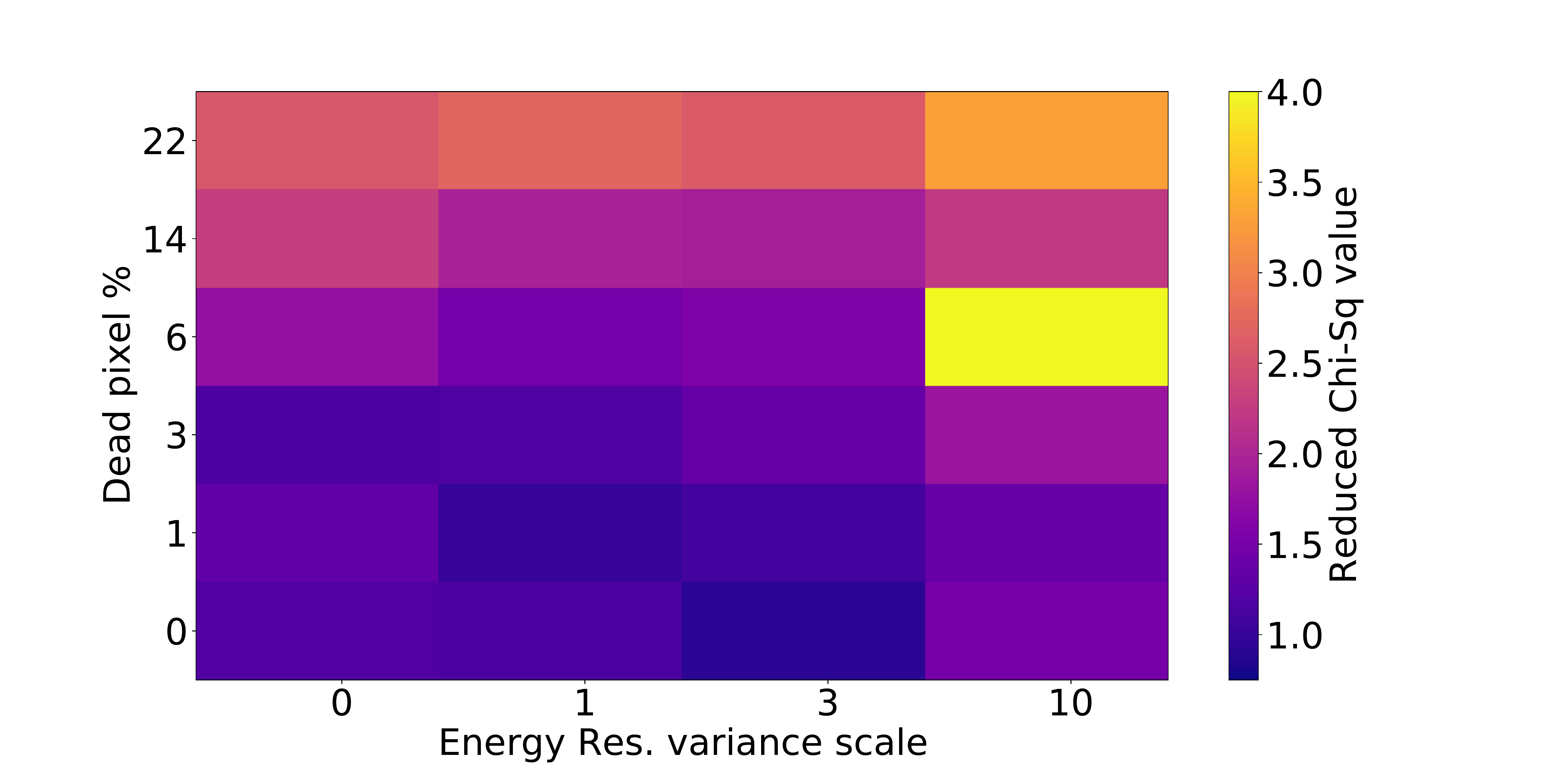}
\includegraphics[width=\columnwidth]{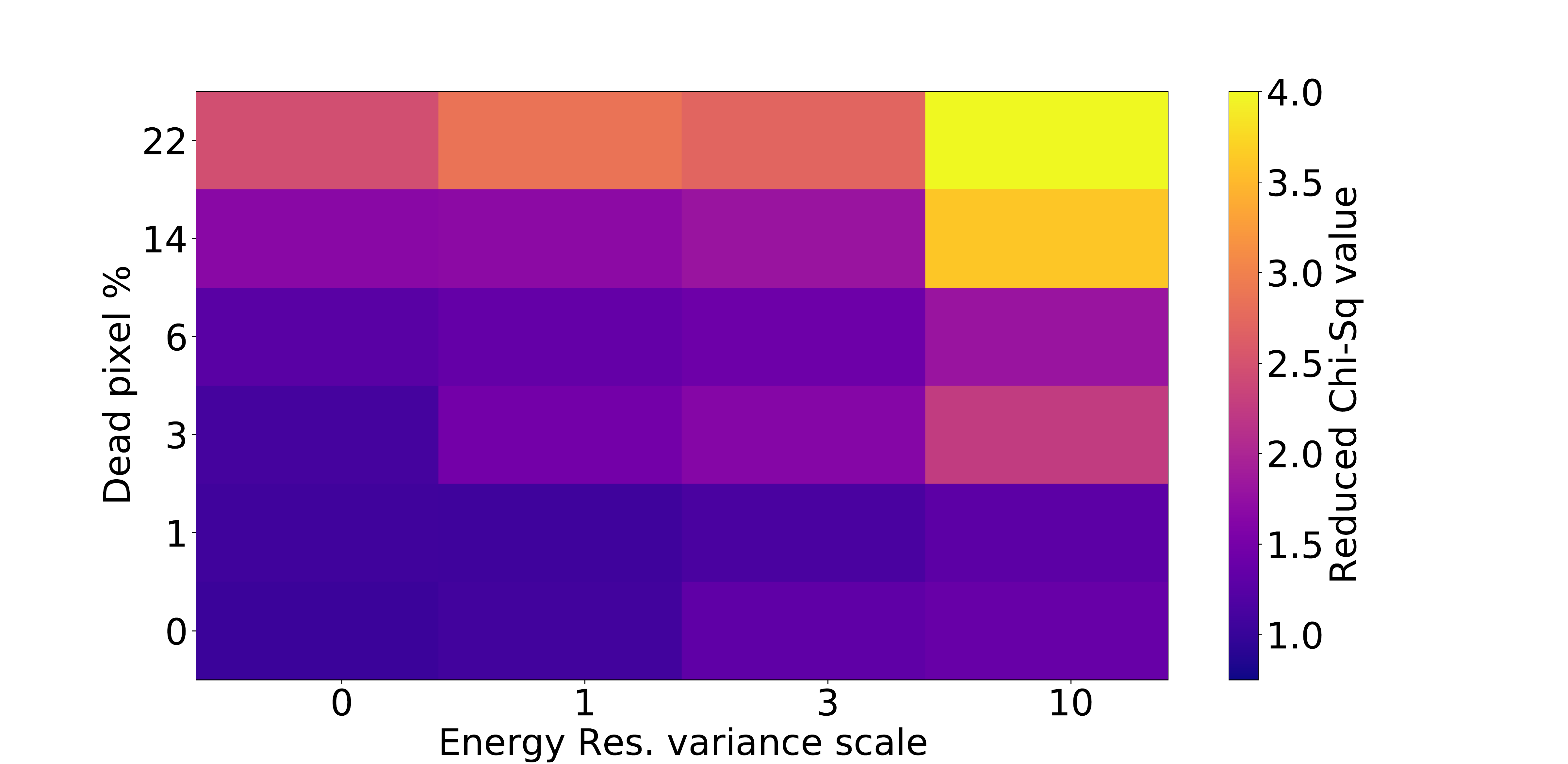}
\includegraphics[width=\columnwidth]{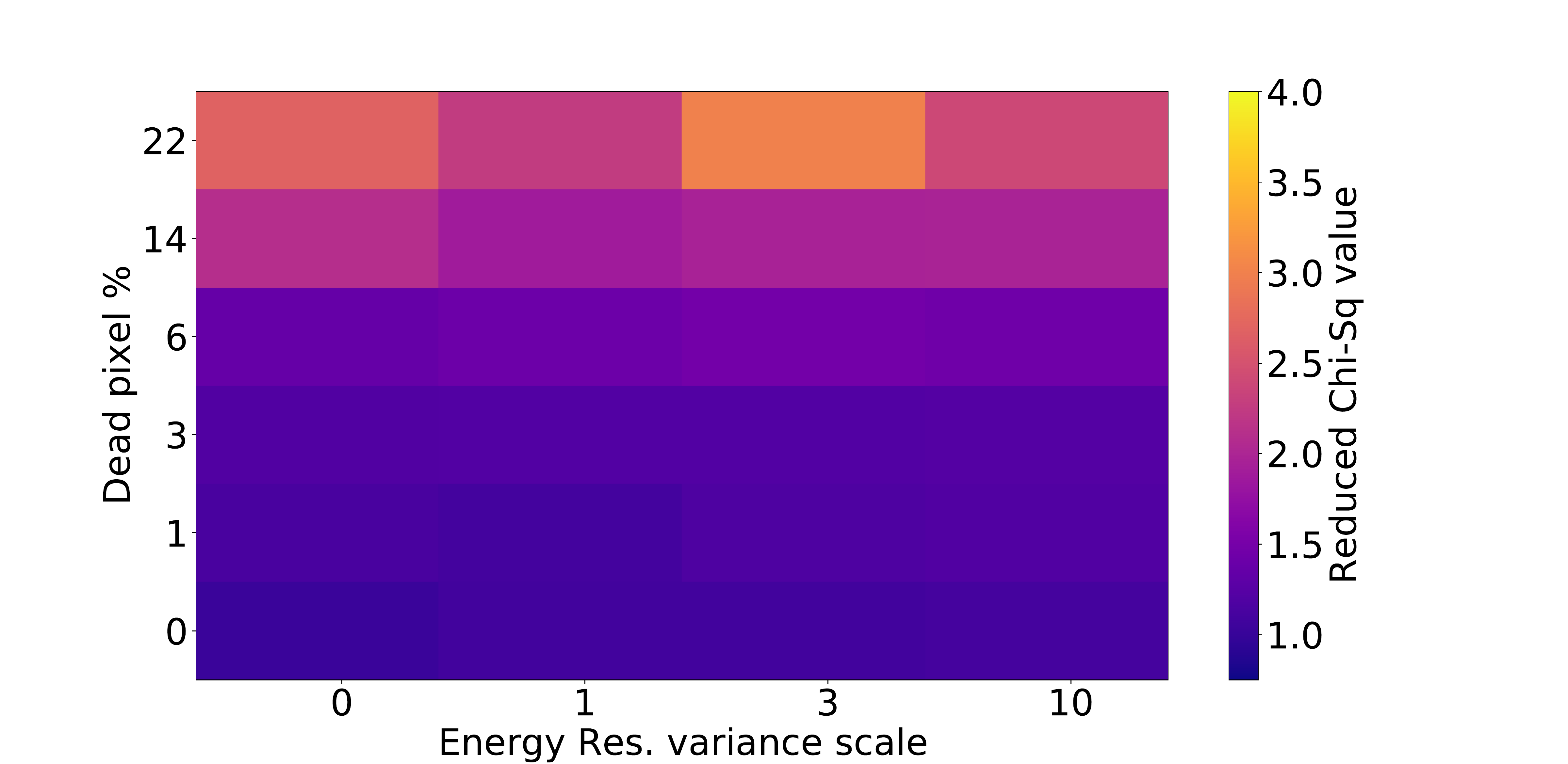}
    \caption{Reduced-Chi Squared results for varying MKID $R_{var}$ and dead pixel percentage, with the Reduced-Chi Squared values representing the colour of the tiles. Simulated using the 8500 spectral resolution KIDSpec setup with $R_{E}=40$, and the two 5000 spectral resolution setups. \textbf{Top Left} is the spectral resolution of 5000 with $R_{E}=25$ setup. \textbf{Top Right} is the spectral resolution of 5000 with $R_{E}=30$ setup. \textbf{Bottom} is the spectral resolution of 8500 with $R_{E}=40$ setup. All grid tiles were averaged over 10 1800s exposures, owing to the photon shot noise.}
    \label{fig: mkid_eff}
\end{figure*}

\subsection{MKID Fabrication Effects}\label{mkid_eff}

To observe the potential impact of MKID effects such as dead pixels and varying $R_{E}$, KSIM can include these effects in a simulation. 

Here the impact of combinations of dead pixels and $R_{var}$ are explored. The dead pixel proportion was allowed to vary up to 22\%, similar to the yield from \cite{Walter2020}, with the dead pixels randomly assigned. Other dead pixel proportions included are; 1\% a yield from a Hawaii 2RG detector currently conducting observations\footnote{\url{https://www.eso.org/sci/facilities/paranal/instruments/crires/doc/CRIRES_User_Manual_P108_Phase2.pdf}}, and 3\% from the yield of a avalanche photodiode array \citep{Atkinson2017DarkArrays}. Other instrument avalanche photodiode arrays have reported yields of 5\%\footnote{\url{https://www.eso.org/observing/dfo/quality/PIONIER/reports/HEALTH/trend_report_BADPIX_HC.html}}. \cite{Meeker2018} showed a $R_{var}$ that occurred in an array of MKIDs for DARKNESS. This variance had an approximately Gaussian form with a standard deviation of 3, centred on 8. To investigate the impact this variance could have on KIDSpec, it is simulated using KSIM. As described in Sec. \ref{pix_gaussian}, $R_{E}$ of the MKIDs were randomly chosen using a normal distribution with varying standard deviations, up to 10. This is a similar ratio to central $R_{E}$ as was observed in \cite{Meeker2018}, owing to the $R_E$ here being higher. It should be noted that these values used for the dead pixels and $R_{var}$ are an upper limit, and are from earlier MKID instruments. Current MKID technology is expected to improve on these values. 

The object used for these simulations was a $m_{R}=20$ mock star-forming galaxy spectrum at redshift $z=2$, from the JAGUAR mock catalogue \citep{Williams2018}, chosen for the wide wavelength range which covers the KIDSpec bandpass in these simulations. An example of this simulation is included in Fig. \ref{fig: jaguar_spec} for the 8500 spectral resolution design. The section between $\approx$1300 to $\approx$1500 nm falls directly in a region of poor atmospheric transmission, which could result in a poorer recreation of the galaxy spectrum. This setup however has omitted this area of atmospheric transmission owing to the optimiser which gives the gap seen in their respective figures. The region of $\approx{350-370}$nm has poor results here also because of the atmospheric transmission reducing quickly from $\approx{400}$nm as can be seen in Fig. \ref{fig: scorer_res}. This transmission resulted in very low numbers of photons reaching the detector stage of the simulation causing this wide spread of flux for this region. The sharp line noise at $\approx{1100}$nm is also the result of low atmospheric transmission and bright sky lines in this region.

To understand how these MKID effects may impact observations, different combinations of these effects were tested for the spectral resolutions of 8500 with $R_{E}=40$, and both 5000 KIDSpec setups. An exposure time of 1800s is used on an 8m diameter telescope. To track KIDSpec performance in observing this spectrum, a Reduced Chi-Squared (RCS) value between the JAGUAR spectrum and the KIDSpec spectrum was used.  
 
Fig. \ref{fig: mkid_eff} contains the results for each combination. All had multiple simulation results averaged, to reduce the impact of photon shot noise. Of the two effects tested, the dead pixels  had the largest impact on the RCS value. For no dead pixels, the $R_{var}$ caused a change of 0.09, 0.34, and 0.30 in the RCS value, from standard deviation values of 0-10 for the $R_{E}=$ 40, 30, and 25 setups respectively. Whereas for no variance, the dead pixels caused a change of 1.66, 1.43, and 1.37 in RCS value from 0-22\% dead pixels, for the $R_{E}=$ 40, 30, and 25 setups respectively. Generally for all plots, a trend of increasing RCS occurs diagonally, following increasing dead pixels and $R_{var}$. Outliers such as the 6\% dead pixels and $R_{E}$ standard deviation of 10 tile, in the $R_{E}=25$ simulations are the results of particularly unfortunate dead or low $R_{E}$ MKIDs. Here these MKIDs have had brighter portions of the JAGUAR spectrum incident on them, and then are dead, have a lower $R_{E}$, or both. 

As mentioned above, for the two 5000 spectral resolution setups, $R_{var}$ had a greater impact than for the higher $R_{E}$ 8500 spectral resolution setup. All $R_E$ standard deviations of 10 had higher RCS values for the $R_{E}=25$ and $30$ setup than the $R_{E}=40$ simulations. This is expected since the value of the standard deviations used across the setup simulations was the same. But this demonstrates that as the $R_{E}$ continues to improve alongside the fabrication, the effect of the $R_{E}$ variation will be mitigated. The dead pixel percentages which follow currently active instruments also did not greatly impact the RCS value. The 1\% dead pixel simulations had average RCS values of only 1.15, 1.11, and 1.21 across the $R_{E}=$ 40, 30, and 25 simulations respectively. For the 3\% dead pixel simulations, the average RCS values were 1.21, 1.55, and 1.26. In particular for the $R_E=40$ simulations these effects did not cause a large increase in RCS.

To mitigate the impact these effects will have on an observation with KIDSpec, any dead pixels, with their locations, and the $R_{E}$s of the MKID array will be known when the MKID array has been fabricated. Additionally because of the lack of read noise, rebinning can be used here with no penalty. For example, if a dead pixel would happen to have a part of a line feature incident on it. With KIDSpec the spectrum could be rebinned, removing that dead bin and only leaving the reduction to the overall SNR. Additionally for both effects, as fabrication methods improve these effects will reduce, further lowering their impact.

\begin{figure*}
\includegraphics[width=18cm]{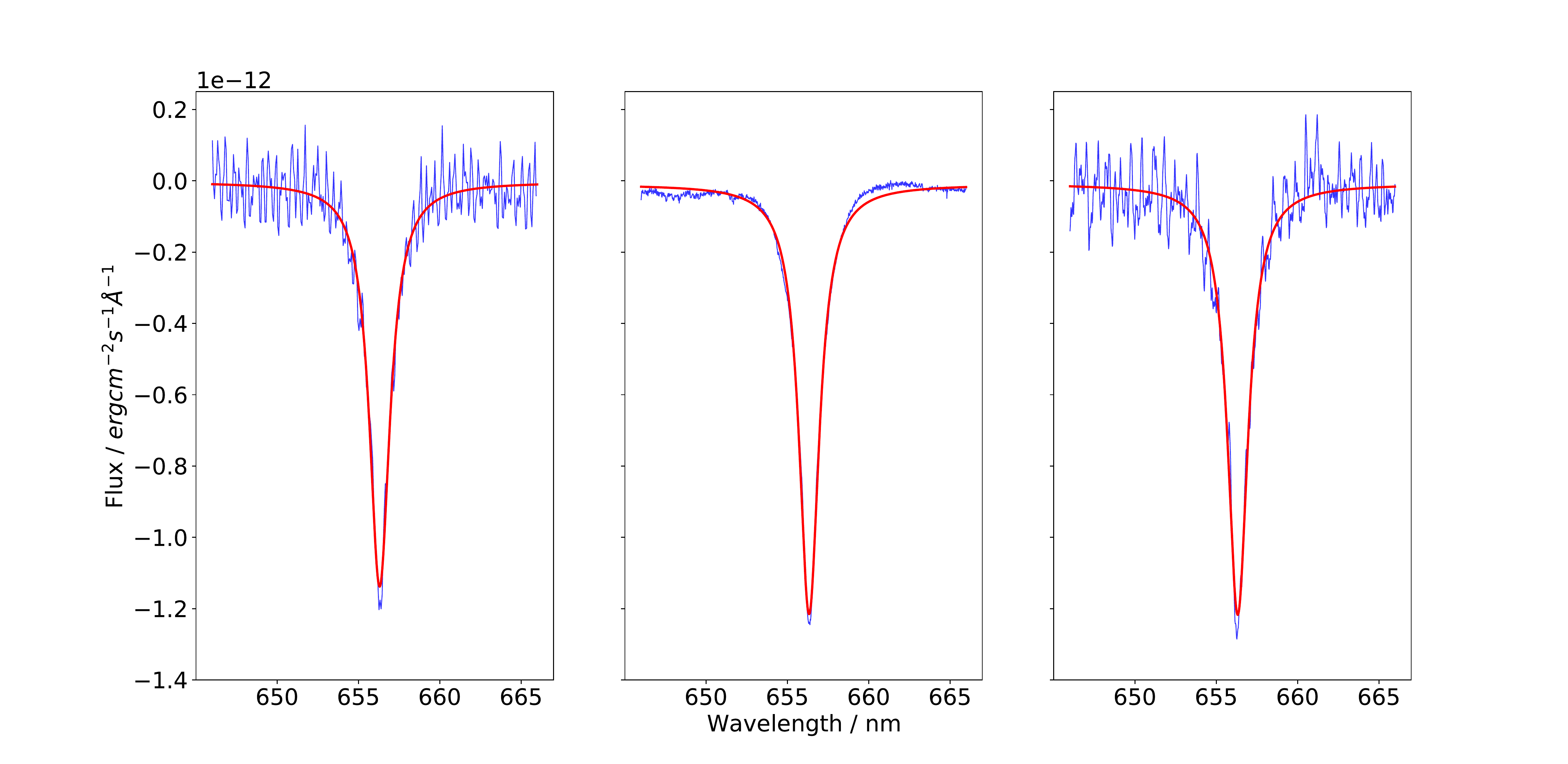}
    \caption{Lorentzian fits of the $H\alpha$ line for the PTS method of KSIM \textbf{(Left)}, the input data spectrum of HD212442 \textbf{(Middle)}, and the Order Gaussian method of KSIM \textbf{(Right)}. The blue spectrum indicates the spectrum used for fitting, and the red line is the resulting fit. These fits respectively gave a FWHM of the $H\alpha$ feature of 1.505$\pm$0.038, 1.483$\pm$0.007, and 1.513$\pm$0.040 nm. Both methods were simulated using the KIDSpec setup with $R_{E}=30$, and a spectral resolution of 5000. The R value comparing the two MKID simulation methods was 0.964. Both methods were simulated using the parameters in Table \ref{table: parameters_runs} and for 60s on a 0.5m telescope.}
    \label{fig: hd212442_fwhm_fit}
\end{figure*}

%----------------------------------------------------------------------
\section{Science Examples}\label{science}

\subsection{Stellar Spectra}\label{stell_spec}

%Spectrum of HD33431 (A0) from ESO's Archive.
%Spectrum of HD212442 (B9) from ESO's Archive.
%Spectrum of HD6229 (G5) from ESO's Archive.

\begin{figure}
\centering
\includegraphics[width=0.5\textwidth]{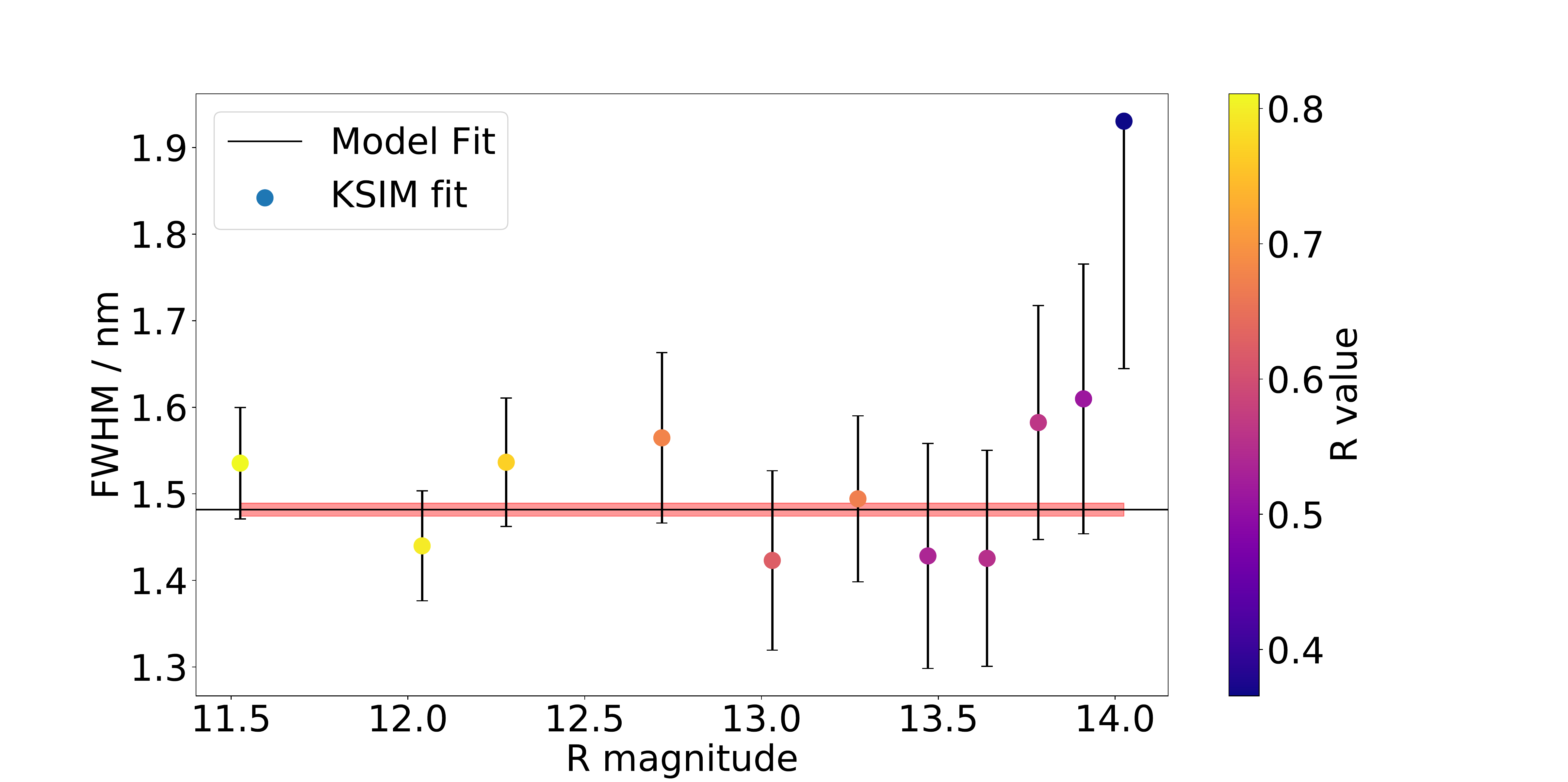}
\caption{Fit results of KSIM simulations of the magnitude reduced spectrum of HD212442, simulated on a 0.5m telescope for 60s using the spectral resolution of 5000 and energy resolution of 30. The bold black horizontal line is the fit to the input spectrum, and the red bar is the error in this fit. The circular points are the KSIM data fit and the errorbars are included for each fit. At $m_{R}\approx{14}$ the R value of the resulting fit reduces to 0.48, below the 0.5 threshold for the fit and data to be strongly correlated.}
\label{fig: hd212442_mag_red_fit}
\end{figure}

\begin{figure*}
\includegraphics[width=18cm]{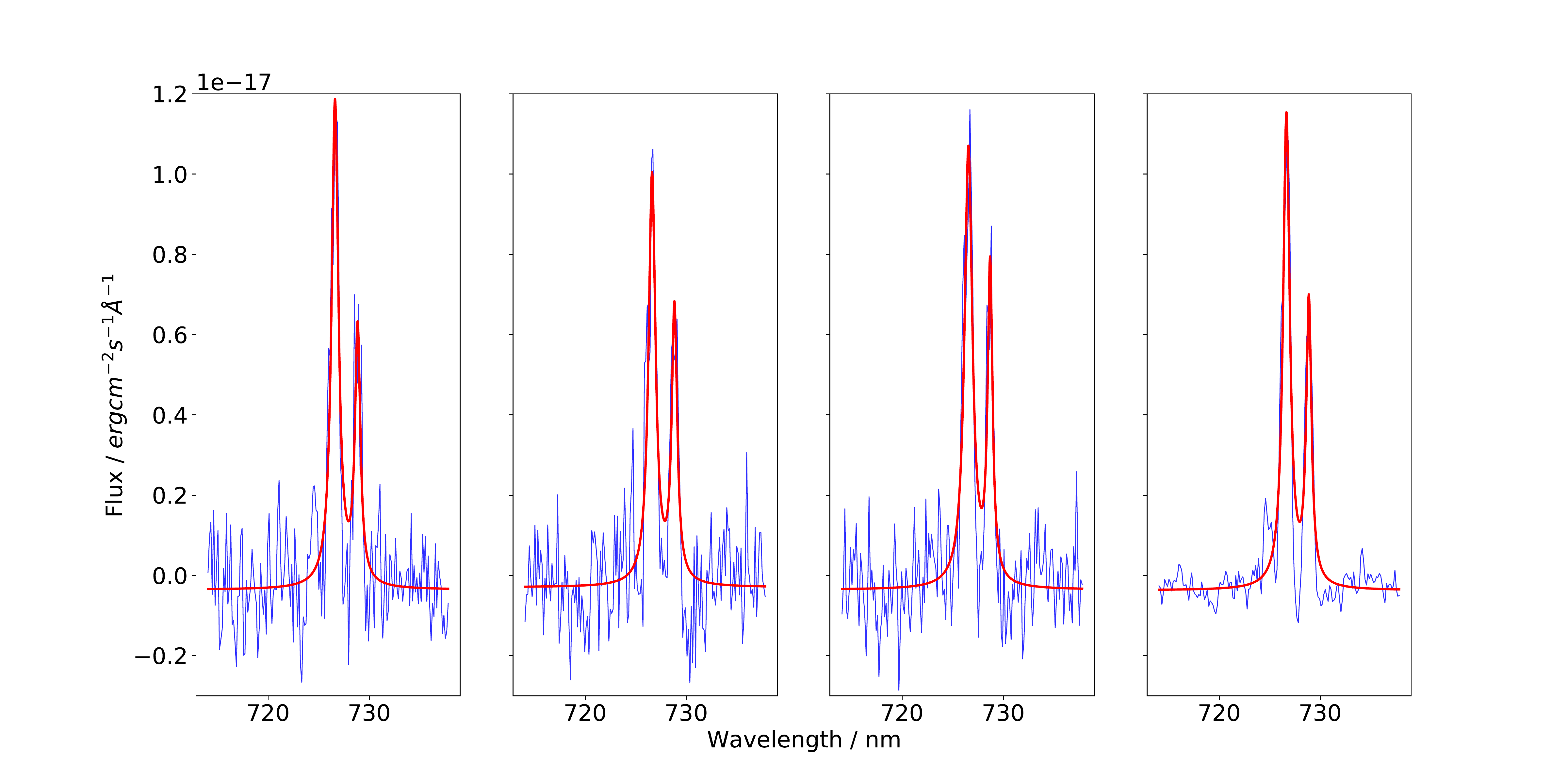}
    \caption{KSIM simulation of SDSS J003948.20+000814.6 ($m_{V}=22$), with an exposure of 180s on an 8m class telescope. Shown are the Lorentzian fit results for the KIDSpec setup with spectral resolution 8500 and $R_{E}=40$ (\textbf{Left}), spectral resolution 5000 and $R_{E}=30$ (\textbf{Middle Left}), spectral resolution 5000 and $R_{E}=25$ (\textbf{Middle Right}), and the original spectral resolution $\approx{2000}$ data spectrum (\textbf{Right}). Included here is the Lorentzian fit of the $H\alpha$ and NII line located at $\approx$726 and 729nm to determine its FWHM. For the $H\alpha$ line this gave 0.792$\pm$0.052 for the input data spectrum from the SDSS DR, and 0.809$\pm$0.084, 0.815$\pm$0.100, and 0.927$\pm$0.096 for the 8500, 5000 ($R_{E}=30$), and 5000 ($R_{E}=25$) spectral resolution setups respectively. For the NII line this gave 0.614$\pm$0.076 for the input data spectrum from the SDSS DR, and 0.634$\pm$0.142, 0.640$\pm$0.130, and 0.568$\pm$0.101 for the 8500, 5000 ($R_{E}=30$), and 5000 ($R_{E}=25$) spectral resolution setups respectively. }
    \label{fig: SDSS_SIM}
\end{figure*}

\begin{figure}
\includegraphics[width=\columnwidth]{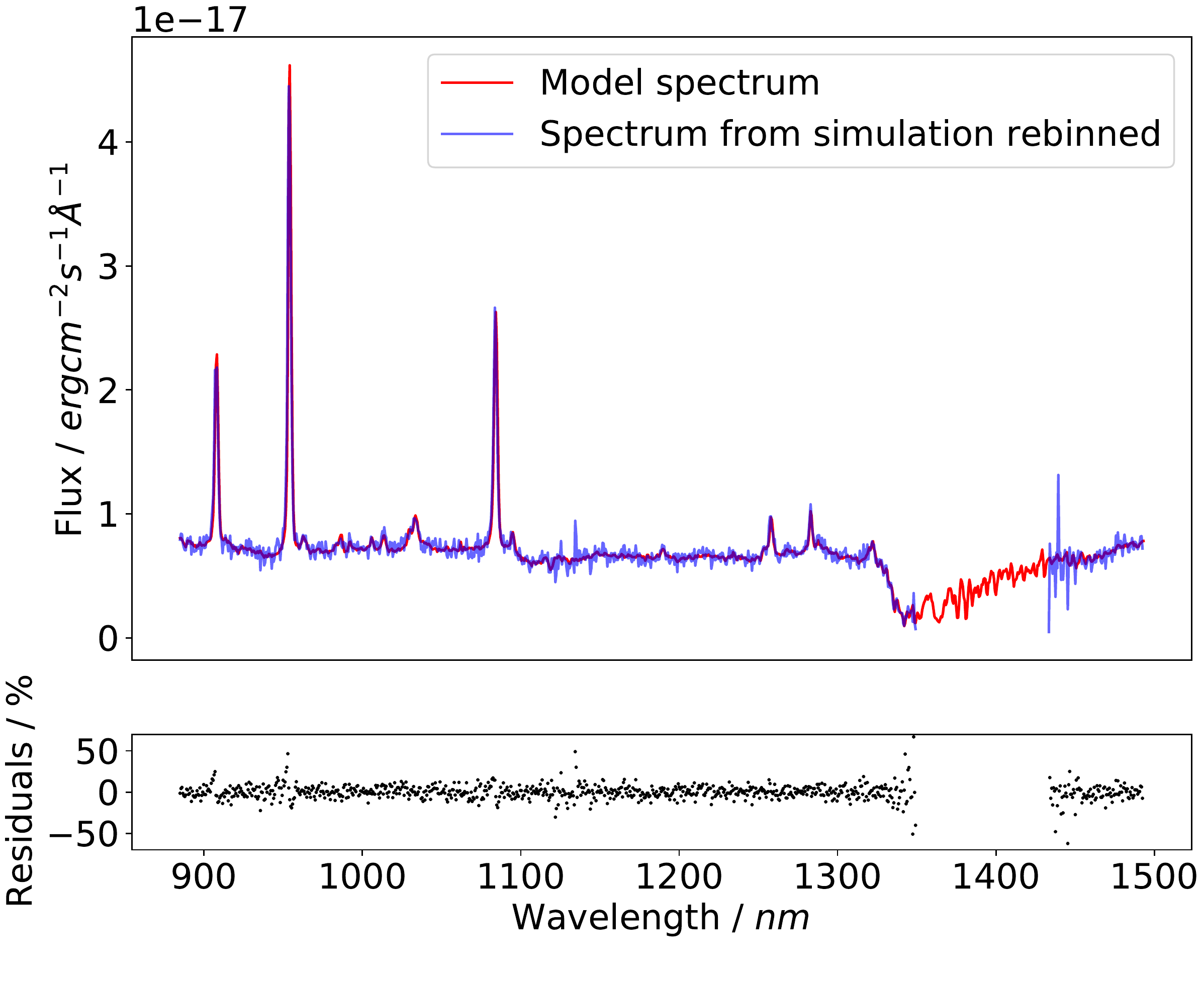}
    \caption{Simulations of Mrk 348 ($m_{R}=21$) for 180s on a 8m telescope using the parameters in Table \ref{table: parameters_runs}. Shown is the result for the KIDSpec setup with a spectral resolution of 8500. The percentage residuals are included in the bottom segment of the Figure. The section between $\approx$1300 to $\approx$1500 nm falls directly in a region of poor atmospheric transmission, which results in a poorer recreation of the galaxy spectrum. This setup however has omitted this area of atmospheric transmission, owing to the optimiser which gives the gap seen in this wavelength range.}
    \label{fig: mrk348_r21}
\end{figure}

HD212442, a B9 type star of $m_{R}=7$, is simulated here to act as a comparison object between the PTS and Order Gaussian methods of simulating the MKID response. The HD212442 spectrum is used from the ESO Archive Library. Parameters used for these simulations are shown in Table \ref{table: parameters_runs}, with exposure time and telescope primary mirror diameter being set to 60 seconds and 0.5 metres respectively. This exposure is chosen here owing to this being the exposure time taken of HD212442 by X-Shooter to gain this data. Here KIDSpec was simulated on a 0.5m diameter telescope, to coincide with the many smaller robotic telescopes currently active in the world, for examples see \cite{Jehin2011}, \cite{Williams2008}, \cite{Steeghs2022}, \cite{Roelfsema2016} and \cite{Boer2003}.  

This comparison of the resulting HD212442 spectrum, between the Order Gaussian and PTS methods gave a R value of 0.964, implying strong positive correlation. Here a value of 1 or -1 implies the two spectra are entirely positively or negatively correlated respectively, greater than 0.5 implies strong positive correlation, and 0 implies entirely no relationship between the two datasets. Additionally the H$\alpha$ feature at $\approx656$nm for both was fitted, as shown in Fig. \ref{fig: hd212442_fwhm_fit}. A Lorentzian fitted to this feature gave a width which, both matched each other and the original data spectrum of HD212442 used, with the Order Gaussian method having a width of 1.513$\pm$0.040 nm and the PTS method had 1.505$\pm$0.038 nm. The original data spectrum of HD212442 had a $H\alpha$ width of 1.483$\pm$0.007 nm, meaning KSIM successfully retrieved this feature across both methods of MKID simulation. From these results we assume the two methods act generally consistently. To observe how faint KIDSpec could still fit this feature for a 60 second exposure on a 0.5m diameter telescope, the magnitude of HD212442 was reduced and simulated using the Order Gaussian method. This was also measured using the Pearson R correlation coefficient value between the KSIM spectrum and the HD212442 spectrum. Fig. \ref{fig: hd212442_mag_red_fit} shows the results of this as HD212442's magnitude was progressively reduced. From Fig. \ref{fig: hd212442_mag_red_fit} KIDSpec could successfully retrieve the $H\alpha$ line to $m_{R}\approx{14}$ on a 0.5m diameter telescope for a 60s exposure. Extending this to a 3600s exposure, the $H\alpha$ line could be retrieved down to $m_{R}\approx{18}$, with a fit of 1.420$\pm$0.141 with a R value of 0.50. 

The theoretical brightness limit for this particular simulation on a 0.5m diameter telescope can also be determined with KSIM. Using the MKID's $\mu$s time resolution and the chosen coincidence rejection time, a maximum number of possible photons per second can be calculated. The same coincidence rejection times discussed in Sec. \ref{PTS}, 200 and 10 $\mu$s, are used here. Considering the $H\alpha$ line again from the HD212442 spectrum, it could be observed to $m_R\approx{0.5}$ for a coincidence rejection time of 200 $\mu$s. For 10 $\mu$s, the line could be observed up to $m_R\approx{-2.4}$, with a range then of down to $m_{R}\approx{18}$, as described above.

\subsection{SDSS J003948.20+000814.6}\label{sdss_gal}

KIDSpec has the potential to be able to observe many galaxy emission lines. As an example of this is the $m_{V}=16$ star forming galaxy SDSS J003948.20+000814.6 was simulated in KSIM. The data for this galaxy was used from the Sloan Digital Sky Survey's (SDSS's) 16th Data Release (DR) \citep{Ahumada2020}. In Table \ref{table: parameters_runs} are the parameters used for the simulation of this target, with telescope diameter and exposure time set to 8m and 60s respectively. These simulations gave an average blaze wavelength SNR of 24.1, 29.2, and 32.8 for the 8500, 5000 ($R_{E}=25$), and 5000 ($R_{E}=30$) spectral resolution setups respectively.

One of the key advantages KIDSpec would have over typical CCD detectors is the lack of read noise and dark current, particularly aiding faint source spectroscopy. To illustrate this, X-Shooter's ETC was used to calculate the predicted SNR for an observation and compare with KIDSpec's. This instrument was chosen because X-Shooter is an optical to NIR spectrograph with comparable spectral resolution. For this comparison a $m_{V}=22$ spectrum of SDSS J003948.20+000814.6 was simulated, for a 180s exposure on an 8m telescope. The mean SNR was taken from the blaze wavelength SNRs from X-Shooter's ETC for this spectrum observation, and the same was done for KSIM's result. Results are shown in Table \ref{table: sim_snrs}. The average SNR for KIDSpec in this comparison was 1.66 and 1.71 for the two spectral resolution of 5000 setups, with $R_{E}$ 25 and 30 respectively. The spectral resolution of 8500 had a average SNR of 1.44. For X-Shooter the average SNRs were 1.10 and 0.83 respectively for the spectral resolution options of 5000 and 8900.

Using its MKIDs, KIDSpec would also be able to flexibly rebin down to lower spectral resolutions, because the detectors do not suffer from read noise. By rebinning down to a spectral resolution of $\approx{2000}$, features can be extracted from faint objects and short exposures. This resolution is the approximate spectral resolution used to originally observe SDSS J003948.20+000814.6. The Lorentzian fitter was used to find the width of the $H\alpha$ and NII lines, located at $\approx$726 and 729nm respectively. For the $H\alpha$ line, this gave 0.792$\pm$0.052 nm for the input data spectrum from the SDSS DR, and 0.809$\pm$0.084, 0.815$\pm$0.100, and 0.927$\pm$0.096 nm for the 8500, 5000 ($R_{E}=30$), and 5000 ($R_{E}=25$) spectral resolution setups respectively. For the NII line, this gave 0.614$\pm$0.076 nm for the input data spectrum from the SDSS DR. KIDSpec achieved 0.634$\pm$0.142, 0.640$\pm$0.130, and 0.568$\pm$0.101 nm for the 8500, 5000 ($R_{E}=30$), and 5000 ($R_{E}=25$) spectral resolution setups respectively. Results are shown in Fig. \ref{fig: SDSS_SIM}. These results demonstrate the potential for KIDSpec to reach higher SNRs with set exposure times, using its lack of read noise and dark current. This is particularly so for these short observations, where KIDSpec still successfully recreates spectral lines.

\subsection{Mrk 348}\label{sey_gal}

A NIR spectrum of Mrk 348 ($m_{R}=14$) was simulated using KSIM, with the parameters used for the simulation in Table \ref{table: parameters_runs}. The data of Mrk 348 shown in \cite{RamosAlmeida2009}, were shared by the author for use in KSIM. Exposure time and telescope diameter was set to 60s and 8m respectively. The average SNR values for these simulations were 82.5, 105, and 102 for the 8500, 5000 ($R_{E}=25$), and 5000 ($R_{E}=30$) spectral resolution setups respectively. To push KIDSpec to more extreme parameters, the $m_{R}$ magnitude of this spectrum was reduced to 21, and simulated with an exposure time of 180s. Results are shown in Fig. \ref{fig: mrk348_r21}. These simulations had average SNR values of 3.63, 4.60, and 4.79 for the 8500, 5000 ($R_{E}=25$), and 5000 ($R_{E}=30$) spectral resolution setups respectively. It is noted that the region of $\approx 1300 - 1500$nm had low flux results for the $R_{E}=25$ setup, which was because this region has poor atmospheric transmission. For this reason the optimiser in Sec. \ref{kidspec_adapt} omitted this region for the other two setups shown in this paper, which can be seen in Fig. \ref{fig: mrk348_r21}. This poor transmission area can also be seen in Fig. \ref{fig: scorer_res}.

From this, as done in Sec. \ref{sdss_gal}, an SNR comparison was done with the X-Shooter ETC. In the NIR KIDSpec is expected to improve over conventional NIR detectors, such as those used for X-Shooter, owing to the MKID's lower noise capabilities described in Sec. \ref{MKIDS}. As in Sec. \ref{sdss_gal}, the setups from the ETC were chosen to match KIDSpec's spectral resolution setups in KSIM. Table \ref{table: sim_snrs} shows the result for this comparison, with X-Shooter having average SNRs of 1.47 and 1.16 for a spectral resolution of 5000 and 8100 respectively. Here, KIDSpec more than doubles these with 3.63, 4.60, and 4.79 for the 8500, 5000 ($R_{E}=25$), and 5000 ($R_{E}=30$) spectral resolution setups respectively.

%From Fig. \ref{fig: mrk348_r21}, lines extending above and below the simulated spectrum (hereafter described as 'bleeds') have appeared, when compared to the top panel of Fig. \ref{fig: SEY_SIM}. Particularly from $\approx1350$nm to 1500nm, this appears to be the result of the atmosphere file used having sections of zero or almost zero in this region, elsewhere this may have been the result of MKID photon misidentfication, explored more in Sec. \ref{mkid_eff}. However, KIDSpec in this simulation still achieved an average SNR of 4.53, almost doubling the X-Shooter like instrument's result of 2.76. This was calculated in the same way as Sec. \ref{sdss_gal}. Full results are shown in Table \ref{table: sim_snrs}.

In Sec. \ref{intro}, the MKID's improved sky subtraction was discussed, namely by utilising a simultaneous sky subtraction with the MKID's time resolving capabilities. To illustrate the impact a simultaneous sky subtraction has in comparison to not, Mrk 348 at $m_{Z}=20$ was simulated with both for 1800s. The simultaneous sky subtraction resulted in an R value of 0.964. The sky spectrum from this exposure was then used for the sky subtraction in a repeated simulation, to replicate a separate sky exposure in an observation. This had an R value of 0.794. When repeated for an exposure of 900s, the simultaneous sky exposure had an R value of 0.960. Whereas the separate sky subtraction had an R value of 0.444, passing the threshold of 0.5 making the result no longer strongly correlated. Additionally, KIDSpec's time resolving capability gives the potential to be able to track variation in the sky lines with photon counting \citep{Mazin2010}. The logistics of using simultaneous sky subtraction could be approached with multiple methods. Some examples would be using a separate MKID array for KIDSpec, or `nodding' the telescope on sky. These aspects could improve KIDSpec's performance particularly in shorter exposures, where KIDSpec has scientific interests.

\begin{table*}
\centering
\caption{This contains SNR results for objects simulated in this paper for KIDSpec, also simulated using X-Shooter's ETC. SNRs below are calculated by taking the mean average of the order blaze wavelength SNRs for those which fall within the object's bandpass. For the instrument comparisons the SNR value included is the average SNR for the simulated object spectrum. The X-S SNR column represents the X-Shooter ETC results. Results included for the JAGUAR mock galaxy are for the simulations without the MKID effects discussed in Sec. \ref{mkid_eff}.}
\label{table: sim_snrs}
\renewcommand\arraystretch{1.5}
\begin{tabular}{@{} c   c  c  c  c  c  c  @{}}
\toprule
Object & Spec. Res. & Energy Res. & Exp. time (s) & Tel. diameter (m) &  \multicolumn{2}{c}{SNR}  \\ 
 & & & & & KIDSpec & X-S \\ \midrule
SDSS J003948.20+000814.6 (V=16) & 8500 & 40 & 60 & 8.0 & 24.1 & 23.8 \\ 
SDSS J003948.20+000814.6 (V=16) & 5000 & 25 & 60 & 8.0 & 29.2 & 30.2  \\
SDSS J003948.20+000814.6 (V=16) & 5000 & 30 & 60 & 8.0 & 32.8 & 30.2  \\
SDSS J003948.20+000814.6 (V=22) & 8500 & 40 & 180 & 8.0 & 1.44 & 0.83  \\
SDSS J003948.20+000814.6 (V=22) & 5000 & 25 & 180 & 8.0 & 1.66 & 1.10  \\
SDSS J003948.20+000814.6 (V=22) & 5000 & 30 & 180 & 8.0 & 1.71 & 1.10  \\
Mrk 348 (R=14) & 8500 & 40 & 60 & 8.0 & 82.5 & 57.8 \\ 
Mrk 348 (R=14) & 5000 & 25 & 60 & 8.0 & 105 & 87.0 \\
Mrk 348 (R=14) & 5000 & 30 & 60 & 8.0 & 102 & 87.0  \\
Mrk 348 (R=21) & 8500 & 40 & 180 & 8.0 & 3.63 & 1.16 \\ 
Mrk 348 (R=21) & 5000 & 25 & 180 & 8.0 & 4.60 & 1.47 \\
Mrk 348 (R=21) & 5000 & 30 & 180 & 8.0 & 4.79 & 1.47 \\
JAGUAR mock galaxy (R=20) & 8500 & 40 & 180 & 8.0 & 3.06 &  1.56  \\ 
JAGUAR mock galaxy (R=20) & 5000 & 25 & 180 & 8.0 & 3.82 & 1.86 \\
JAGUAR mock galaxy (R=20) & 5000 & 30 & 180 & 8.0 & 4.00 & 1.86  \\
\bottomrule
\end{tabular}
\end{table*}

%----------------------------------------------------------------------
\section{Discussion}\label{discuss}

As shown in this paper KIDSpec will have the ability to observe a variety of astronomical targets, and contribute to a wide range of science areas. The spectra of stars carry important information about them; abundance of elements, effective temperature, and gravity for example \citep{Szczerba2020}. Galaxy observations also present opportunities for many different science areas. High redshift star forming galaxies and their emission lines can provide information on the stages of evolution of the universe, such as the epoch of reionisation \citep{Onodera2020}. Emission line galaxies at medium to high redshifts can aid in testing the cosmological dark energy model and constrain it \citep{Gao2020}. The central phase of reionisation of hydrogen in the universe can be investigated using Ly$\alpha$ emitter galaxies, because of the resonant scattering of Ly$\alpha$ photons being responsive to the neutral hydrogen in the intergalactic medium \citep{Yang2019}. KIDSpec's relevant features for this are that MKIDs do not suffer from readout noise, and KIDSpec would have superb cosmic ray removal owing to the MKID's time resolution \citep{Mazin2013}. In an MKID, the cosmic ray event lasts a few hundred microseconds and can be easily removed from the phase time-stream because of its amplitude. This will result in the loss of a small fraction of an exposure. In a semiconductor detector, all of the charge from the cosmic ray is stored until the readout, affecting the whole exposure time \citep{OBrien2020}.

Also as discussed in Sec. \ref{sdss_gal}, because of the MKID's lack of readout noise, the resulting spectrum can be rebinned flexibly to a lower spectral resolution.

Active galactic nuclei can be spectrally identified using observed emission lines and from this, information on their stellar populations can be gained. NIR spectroscopy is well suited to this as it is in a less extinct wavelength range when compared to the optical. NIR spectroscopy is not dominated by nonstellar emission, allowing for the study of the stellar content of the galaxy, and the search for signatures of intense star formation \citep{RamosAlmeida2009}.

Other science areas of interest to KIDSpec include spectrally time resolved studies such as orbits of compact binaries. These observations are currently limited, especially ahead of the launch of the Laser Interferometer Space Antenna (LISA) \citep{Amaro-Seoane2017}. LISA will aim to detect tens of thousands of these systems \citep{Burdge2019}. For short period binaries, $\approx15$ minutes, one limitation of CCDs is their read noise. A sufficiently long exposure must be taken to reduce the read noise's impact, which limits the number of phase bins of the binary's period to constrain its parameters. The effect of this is an increase in error on the parameters, possibly even to similar values to the found feature \citep{Burdge2019}. Additionally, smearing of spectral features for systems with high radial velocity make long exposures also unsuitable, as the spectral line would not be resolvable. MKIDs however, are not limited by exposure time owing to a lack of read noise, and therefore can use more phase bins, i.e. shorter exposures, and hence result in more constrained parameters. The LRIS instrument \citep{Epps1995} was used for spectroscopic observations of the system in \cite{Burdge2019}. LRIS can have a spectral resolution between 300 and 5000 with a bandpass of 320 to 1000 nm, making the 5000 spectral resolution KIDSpec setup described here an appropriate option. The magnitudes for 30s exposures with an SNR $\geq10$, calculated in Sec. \ref{kidspec_adapt}, at a spectral resolution of 5000 approach the $\approx{19}$ AB magnitude of the system observed in \cite{Burdge2019}. Using KIDSpec would allow for more phase bins during an observation of these systems, reducing the final error on the parameters. Additionally KIDSpec could take a continuous exposure and the observation can be separated afterwards because of the time resolving capabilities of the MKIDs. This makes KIDSpec suited to observing short period binaries, such as the systems LISA will observe.

Another field of interest is fast time variability systems such as pulsars, magnetic white dwarfs, and other binary systems \citep{Mazin2004}. For pulsars and magnetic white dwarfs, optical data alongside simultaneous radio and gamma ray observations, will aid testing of high-energy emission models \citep{Mazin2010}. KIDSpec will also be able to constrain parameters of various types of binaries \citep{OBrien2020}. In the Bowen region of 463 to 466 nm, there are narrow high excitation emission lines, and these lines appear to move during the period of the binary system. Using the movement of these lines the semi amplitude of the radial velocity and mass function can be constrained \citep{Cornelisse2008}. Discussed here are just a selection of KIDSpec's science cases, which continues to grow.

%-----------------------------------------------------------------------
\section{Conclusions}\label{conclusions}
A new simulation tool for the proposed medium resolution spectrograph instrument KIDSpec has been described, with astrophysical objects simulated. KIDSpec aims to set itself apart with its use of MKIDs. These superconductor detectors bring several benefits, including time resolution, no read noise, no dark current, and excellent cosmic ray mitigation. The intrinsic energy resolving capabilities of MKIDs mean a cross-disperser is not required, as MKIDs can separate the orders themselves, simplifying the optical layout of KIDSpec. Resolving the orders represents one of the greatest challenges for KIDSpec. Its capabilities will be limited by the $R_{E}$ of its MKIDs, which limits how many orders can be observed by a single MKID. The benefits of MKIDs makes KIDSpec suitable for a growing list of science, including faint source spectroscopy and short period binaries, owing to the MKID's lack of read noise and dark current, and time resolution.

The KIDSpec Simulator (KSIM) has been developed to evaluate KIDSpec's performance, and is now in a position to test more KIDSpec parameters and new science cases. By simulating the atmosphere, telescope, grating, MKIDs, and other aspects, KSIM simulates how KIDSpec could observe a given object. Through the Order Gaussian or PTS method, KSIM can now aid the future development of the KIDSpec instrument. Using KSIM's variety of input parameters, observing scenarios, and objects can be simulated. Simulated in this work were various stars and galaxies, including comparisons with X-Shooter's ETC. It was found that for short observations involving fainter objects, KIDSpec potentially doubles the SNR of the same simulated observation with X-Shooter.

MKID fabrication effects were also simulated to observe the impact they may have on observations. The dead pixels most affected the recreation of the object spectra, while $R_{var}$ had a lower observed impact on the RCS value. Additionally, these fabrication errors will be further mitigated as the technology and fabrication continues to improve, and with testing KIDSpec's MKID array after fabrication to find the locations of any dead pixels. Multiple designs were also simulated to predict limiting magnitudes for various KIDSpecs, which found it would have comparable or fainter limiting magnitudes than X-Shooter. This work demonstrates KSIM's ability and flexibility to test a variety of science cases and KIDSpec designs, to progress the realisation of KIDSpec as a constructed instrument. To further grow the list of science cases for KIDSpec objects can be simulated using KSIM on request, or the simulation tool can also be obtained online.

\section*{Acknowledgements}

For the purpose of open access, the authors have applied a Creative Commons Attribution (CC BY) licence to any Author Accepted Manuscript version arising from this submission. The authors would like to thank the anonymous referees for improving this paper with their feedback, Dr. C. Ramos Almeida for sharing spectra to be run through KSIM to help demonstrate KIDSpec's potential, and Prof. T. Marsh for helpful discussions relating to observing cataclysmic variables and other binary systems. VBH and KOB are supported by the Science and Technology Facilities Council (STFC) under grant numbers ST/T506047/1, ST/T002433/1, and ST/X003280/1. This project has received funding from the European Union's Horizon 2020 research and innovation programme under grant agreement No 730890.

%%%%%%%%%%%%%%%%%%%%%%%%%%%%%%%%%%%%%%%%%%%%%%%%%%
\section*{Data Availability}

KSIM can be obtained from the GitHub repository located at \textit{https://github.com/BVH1979/KIDSpec-Simulator-KSIM}.



\bibliographystyle{rasti}
\bibliography{references2}
%\bibliography{references}% if your bibtex file is called example.bib

%\begin{thebibliography}{99}
%\end{thebibliography}
%%%%%%%%%%%%%%%%%%%%%%%%%%%%%%%%%%%%%%%%%%%%%%%%%

%%%%%%%%%%%%%%%%% APPENDICES %%%%%%%%%%%%%%%%%%%%%

\appendix

\section{KSIM Parameter List}\label{append_a}
Table \ref{tab:ksim_params} contains all variable parameters for KSIM, with units and formats.

\clearpage
\onecolumn
\renewcommand{\arraystretch}{1.3}
\begin{longtable}{p{5cm}|p{12cm}}
\caption{All parameters which can be altered for a astronomical target object observation simulation using KSIM. Where appropriate, a requirement or range for the value of the parameter is also included.}\label{tab:ksim_params}\\
Parameter & Description \\ \hline
\textit{object\_name} & Name of astronomical target object being simulated within KSIM. \\ 
\textit{object\_file} & Name of file containing spectrum of astronomical target object, structured in the form of two columns with wavelength (nm) and flux ($ergcm^{-2}s^{-1}\text{\r{A}}^{-1}$). \\
\textit{binstep} & The size in nm of the bins in the Object File spectrum.\\
\textit{mirr\_diam} & Diameter in cm of the primary mirror of the telescope.  \\
\textit{central\_obscuration} & Percentage obscuration to the primary mirror of the telescope. \\
\textit{seeing} & Value of the atmospheric seeing, in arcseconds.  \\
\textit{exposure\_t} & Exposure time of simulated observation in seconds. \\
\textit{tele\_file} & Text file containing two columns, wavelength in nm and percentage reflectance of telescope mirror material. \\
\textit{lambda\_low\_val} & Minimum wavelength for simulated KIDSpec bandpass. Minimum value of 350nm and maximum value of 2999nm. \\
\textit{lambda\_high\_val} & Maximum wavelength for simulated KIDspec bandpass. Minimum value of 351nm and maximum value of 3000nm. \\
\textit{n\_pixels} & Number of MKID pixels in linear array for KIDSpec. Minimum value greater than zero.\\
\textit{alpha\_val} & Incidence angle of incoming light to grating in degrees. \\
\textit{phi\_val} & Reflected central angle of incoming light to grating in degrees. \\
\textit{refl\_deg} & Reflected angle range of incoming light passed to MKIDs in degrees.       \\                                                          
\textit{grooves} & Number of grooves on grating per mm.            \\                           
\textit{norders} & Number of grating orders to test for incoming wavelengths. \\ 
\textit{number\_optical\_surfaces} & Number of optical surfaces in KIDSpec instrument between primary mirror and MKIDs. The GEMINI silver mirrors reflectance is used here.\\
\textit{folder\_dir} & Folder path where all other files can be found and where results are saved to.\\ 
\textit{fudicial\_energy\_res} & Energy resolution used to calculate energy resolution at all other wavelengths. \\    
\textit{fudicial\_wavelength} & Wavelength used to calculate energy resolution at all other wavelengths. \\               
\textit{coincidence\_rejection\_time} & The coincidence rejection time, in $\mu$s, used for MKID saturation calculations for both the PTS and Order Gaussian methods. \\
\textit{raw\_sky\_file} & FITS file containing the sky background, can be generated using ESO SKYCALC.\\       
\textit{slit\_width} & Width of slit in arcseconds. \\                     
\textit{pixel\_fov} & FoV of MKID pixels in arcseconds. \\           
\textit{off\_centre} & Sets the distance target object is from the centre of the slit in arcseconds. Can be set to zero or greater.\\  
\textit{airmass} & Airmass of atmosphere. \\                    
\textit{dead\_pixel\_perc} & Percentage of MKIDs which are considered dead. Value can be set in the range 0-100.\\        
\textit{R\_E\_spread} & Standard deviation value of normal distribution used to generate spread of $R_{E}$. Can be set to zero or greater.\\               
\textit{redshift} & Desired redshift of target object.\\ 
\textit{redshift\_orig} & Original redshift of target object.\\
\textit{mag\_reduce} & Factor which reduces incoming flux from simulated target. Can be set to <1 for an increase in flux.\\       
\textit{generate\_sky\_seeing\_eff} & Generates transmission file, containing transmission of sky spectrum though slit. \\      
\textit{sky\_seeing\_eff\_file\_save\_or\_load} & Name of sky seeing transmission file to either save or load.\\      
\textit{generate\_model\_seeing\_eff} & Generates transmission file, containing transmission of target object spectrum though slit\\                               
\textit{model\_seeing\_eff\_file\_save\_or\_load} & Name of target object seeing transmission file to either save or load.\\    
\textit{generate\_additional\_plots} & Plots additional steps throughout KSIM, including photon spectra at various stages such as atmosphere, telescope, and grating orders.\\                
\textit{generate\_standard\_star\_factors} & Generates standard star spectral weights as described in Sec. \ref{sim_out}.\\          
\textit{stand\_star\_run\_filename\_details} & Name of standard star spectral weights to either save or load.\\ 
\textit{fwhm\_fitter} & Option to use a Lorentzian shape fitter for spectral features, up to two features at once.\\
\textit{fwhm\_fitter\_central\_wavelength} & Central wavelength of a Lorentzian shaped line.\\
\textit{fwhm\_fitter\_central\_wavelength\_2} & Central wavelength of a second Lorentzian shaped line.\\  
\textit{double\_fitter} & If two lines are to be fitted then this is set to True.\\
\textit{continuum\_removal\_use\_polynomial} & If True a polynomial will be fitted to the spectrum to remove the spectrum continuum. If False a linear fit is used.\\    
\textit{reset\_R\_E\_spread\_array} & When True generates new energy resolution spreads. \\         
\textit{reset\_dead\_pixel\_array} & When True generates dead pixel spreads. \\            
\hline
\end{longtable}
%\clearpage
%\twocolumn
%%%%%%%%%%%%%%%%%%%%%%%%%%%%%%%%%%%%%%%%%%%%%%%%%%

% Don't change these lines
\bsp	% typesetting comment
\label{lastpage}
\end{document}